\def\spose#1{\hbox to 0pt{#1\hss}}
\def\ltsimm{\mathrel{\spose{\lower 3pt\hbox{$\sim$}}
        \raise 2.0pt\hbox{$<$}}}
\def\gtsimm{\mathrel{\spose{\lower 3pt\hbox{$\sim$}}
        \raise 2.0pt\hbox{$>$}}}

\documentclass[useAMS,usenatbib]{mn2e}
\usepackage{graphicx}
\usepackage{multirow}
\usepackage{amsmath}
\usepackage{amssymb}

\title[Feedback from Winds and Supernovae in Massive Stellar Clusters. II: X-Ray Emission]{Feedback from Winds and Supernovae in Massive Stellar Clusters. II: X-Ray Emission}
\author[H.Rogers and J.M.Pittard]{H.Rogers and J.M.Pittard\\
School of Physics and Astronomy, The University of Leeds, Leeds, LS2 9JT}

\date{Received 17 February 2014.  Accepted 31 March 2014}

\pagerange{\pageref{firstpage}--\pageref{lastpage}} \pubyear{2014}

\def\LaTeX{L\kern-.36em\raise.3ex\hbox{a}\kern-.15em
    T\kern-.1667em\lower.7ex\hbox{E}\kern-.125emX}

\begin{document}

\maketitle

\label{firstpage}

\begin{abstract}
The X-ray emission from a simulated massive stellar cluster is investigated.  The emission is calculated from a 3D hydrodynamical model which incorporates the mechanical feedback from the stellar winds of 3 O-stars embedded in a giant molecular cloud (GMC) clump containing 3240\,M$_{\odot}$ of molecular material within a 4 pc radius.  A simple prescription for the evolution of the stars is used, with the first supernova explosion at t\,=\,4.4\,Myrs.  We find that the presence of the GMC clump causes short-lived attenuation effects on the X-ray emission of the cluster.  However, once most of the material has been ablated away by the winds the remaining dense clumps do not have a noticable effect on the attenuation compared with the assumed interstellar medium (ISM) column.  We determine the evolution of the cluster X-ray luminosity, L$_X$, and spectra, and generate synthetic images.  The intrinsic X-ray luminosity drops from nearly 10$^{34}$\,ergs\,s$^{-1}$ while the winds are `bottled up', to a near constant value of 1.7$\times 10^{32}\rm\,ergs\,s^{-1}$ between t\,=\,1--4\,Myrs.  L$_X$ reduces slightly during each star's red supergiant (RSG) stage due to the depressurization of the hot gas.  However, L$_X$ increases to $\approx\,10^{34}\rm\,ergs\,s^{-1}$ during each star's Wolf-Rayet (WR) stage.  The X-ray luminosity is enhanced by 2-3 orders of magnitude to $\sim\,10^{37}\rm\,ergs\,s^{-1}$ for at least 4600 yrs after each supernova (SN) explosion, at which time the blast wave leaves the grid and the X-ray luminosity drops.  The X-ray luminosity of our simulation is generally considerably fainter than predicted from spherically-symmetric bubble models, due to the leakage of hot gas material through gaps in the outer shell.  This process reduces the pressure within our simulation and thus the X-ray emission.  However, the X-ray luminosities and temperatures which we obtain are comparable to similarly powerful massive young clusters.
\end{abstract}


\section{Introduction}
Massive stars have a profound affect on their natal environment.  They have strong ionizing radiation fields which create HII regions, and their powerful winds sweep up surrounding material creating wind-blown shells and cavities.  Their supernovae (SNe) chemically enrich the interstellar medium (ISM) and help to sustain turbulence within it.  Thus the presence of massive stars in stellar clusters has implications for future generations of star formation.  The dispersal and destruction of molecular material by the winds and ionzing radiation may inhibit further star formation within that region.  Conversely, compression of material by winds and shocks may also trigger new star formation \citep{Koenig12} and new cluster formation \citep{Beuther08,Gray11}.  

Many high-mass star-forming regions are observed to contain diffuse thermal X-ray emission, which requires high temperature plasma. It has long been recognized that the fast winds of individual massive stars create high pressure and high temperature bubbles \citep[e.g.][]{Dyson72,Castor75,Weaver77}. In large clusters containing many early-type stars the individual stellar winds may combine, collectively creating a so-called cluster wind \citep[e.g.][]{Chevalier85,Canto00,Stevens03}. 

In many cases the observed diffuse emission from massive-star forming regions (MSFRs) is relatively soft. For instance, in M17 and the Rosette nebula the characteristic temperature $kT < 1 $\,keV \citep{Townsley03}. The X-ray emitting plasma in the Extended Orion nebula is similarly cool \citep{Gudel08}. However, in other MSFRs the characteristic temperature is considerably higher. For instance, the diffuse thermal X-ray emission from NGC\,3603 \citep{Moffat02}, the core of the Arches cluster \citep{Wang06}, and the Quintuplet cluster \citep{Law04} has $kT > 2$\,keV. 

In some clusters the diffuse X-ray emission may be predominantly non-thermal: e.g. NGC\,6334 \citep{Ezoe06a}, RCW\,38 \citep{Wolk02}, the Arches cluster \citep{Yusef-Zadeh02,Wang06} and ON2 \citep{Oskinova10}. This requires non-thermal particles, which also occur when high speed flows and strong shocks are present. \citet{Townsley11} argue that the non-thermal emission detected from the NGC\,3576~OB~association may arise from both a pulsar wind nebula and a cavity supernova. A caveat to some of these works is that the diffuse X-ray emission from clusters which are at larger distances is more likely to suffer contributions from unresolved point sources. A summary of X-ray studies of young stellar clusters is presented by \citet{Damiani10}. 

Early studies indicated that the detection of diffuse X-ray emission in stellar clusters required the presence of stars earlier than O6 \citep{Townsley03}, although an exception, the O7-powered Hourglass nebula \citep{Rauw02}, was known. \citet{Ezoe06b} have since discovered diffuse X-ray emission from NGC\,2024 (the Flame nebula), which contains only late O- to early B-type stars. On the other hand, it is curious that diffuse X-ray emission has yet to be detected from some very massive stellar clusters where very early O-type stars are present, such as 
Trumpler\,16 \citep{Wolk11}. 

Other work has indicated that the temperature of the diffuse plasma may be correlated with how embedded it is, as measured by the column density
\citep{Ezoe06a}. Clusters with high temperature plasma appear to have $N_{\rm H}>5\times10^{21}\,{\rm cm^{-2}}$, while clusters with cooler plasma have less absorption. \citet{Wolk08} suggest that while the stellar winds are bottled up the shocked gas remains maximally heated, but subsequent leakage and the resulting adiabatic cooling of the gas causes the gas temperature to drop.

Unfortunately, past comparisons of X-ray observations with theory have had mixed success.  Many works on MSFRs simply compare the observed X-ray luminosity against the mechanical wind power of the stars, or the thermal energy of the plasma against an estimate of the time-integrated energy input of the winds \citep[e.g.][]{Townsley03,Ezoe06b,Gudel08}. The efficiency of the conversion of mechanical energy to radiation is then found to range from $~10^{-4}$ to 0.1. This, and the estimated mass of the X-ray emitting gas, indicates that in many cases the winds are not completely confined and that hot plasma must flow into the wider environment. This conclusion is reinforced by the fact that the application of completely confined wind-blown-bubble models often leads to a significant overprediction of the X-ray luminosity \citep[e.g.][]{Rauw02,Dunne03,Harper-Clark09}. 

Other works have compared the X-ray luminosity and the surface brightness profile of the diffuse emission to the predictions of cluster-wind models. In their analysis of the Arches and Quintuplet clusters, \citet{Wang06} found that the radial intensity profiles of the diffuse emission were more extended than theoretical predictions.%

\citet{Harper-Clark09} recently determined that the observed diffuse X-ray emission from the Carina Nebula was 60 times too faint compared to predictions from the \citet{Castor75} model, and 10 times too luminous compared to the \citet{Chevalier85} model.  This led \citeauthor{Harper-Clark09} to develop a third model whereby density variations in the ISM surrounding the cluster causes gaps in the swept-up shell, through which some of the high pressure gas in the bubble interior can leak.  Their new model predicts a lower pressure within the bubble than the \citeauthor{Castor75} model, as the wind material is not completely confined, and also a lower X-ray luminosity.  Consequently, it is more consistent with observations.  However, since the covering fraction of the shell is a free parameter this model suffers from a lack of predictive power.

In young MSFRs in which there has not yet been time for any massive star to explode as a supernova, the diffuse X-ray emission must result from the action of stellar winds. However, in older clusters where some of the massive stars have exploded one still might not detect any signature of a SN explosion because the effect of a SNR on the thermal properties of the hot cluster gas is likely to be relatively short-lived. This time scale is generally believed to be $\sim 10^{4}$\,yr \citep[e.g.][]{Kavanagh11}. For this reason, most studies of stellar clusters prefer a wind based explanation for the diffuse X-ray emission, though \citet{Ezoe09} favour a recent SN explosion in their study of the Eastern Tip of the Carina nebula.  A distinction exists between individual stellar clusters, and larger scale regions of star formation which create superbubbles where multiple cavity supernovae are believed to be responsible for the diffuse emission \citep[such as those of 30\,Doradus, e.g.][]{Chu90, Townsley11}.

Given the challenges of interpreting such complex environments as MSFRs, and the highly idealized models of most theoretical and modelling work,
in this paper the hydrodynamical models of stellar wind and supernova feedback in an inhomogeneous environment outlined in \citet{Rogers13} (henceforth referred to as Paper I) are used as a basis to simulate the resulting X-ray emission from such regions. Of great interest are the X-ray luminosity and spectrum, and their temporal variation as the stars in the simulation cycle through various evolutionary stages, including main sequence, red supergiant, Wolf-Rayet and supernova. In Section~\ref{sec:xray_model_details} the details of the model and the method of calculating the X-ray emission and absorption are discussed.  The results are presented in Section~\ref{sec:xray_results}.  Comparisons to numerical models and observations are made in Sections~\ref{sec:comp_theory} and~\ref{sec:comp_obs} respectively.  Section~\ref{sec:xray_conclusions} summarises and concludes this work.


\section{Simulations}
\label{sec:xray_model_details}
\subsection{The Numerical Model}
The X-ray calculations in this paper are based on the 3D hydrodynamical model described in Paper I.  The simulations were performed using the hydrodynamical code ARWEN, which uses a piecewise parabolic interpolation and characteristic tracing to obtain the time-averaged fluid variables at each zone interface.  An iterative Riemann solver is used to determine the time-averaged fluxes and solve the equations of hydrodynamics (see Paper I for more details).  The simulations consist of three massive O stars which represent the main sources of mechanical feedback in a massive star forming region contained within an inhomogeneous GMC clump of radius 4\,pc and mass 3240\,M$_{\odot}$.  The medium surrounding this clump is homogeneous, with a density of 3.33$\times10^{-25}\rm\,g\,cm^{-3}$ ($n_H \approx n_e \approx 0.2\rm\,cm^{-3}$) and a temperature of 8000\,K.  The simulations were performed on a 512$^3$ grid with free outflow boundary conditions.  The total simulation volume covers a cubic region of $\pm$\,16\,pc centered on the GMC clump.  The cluster wind is injected as thermal energy within a radius of 6 cells (0.375\,pc).  

The evolution of the three stars is treated simplistically as three distinct phases - the Main Sequence (MS), Red Supergiant (RSG) and Wolf-Rayet (WR) phases.  The details of the stellar cluster are summarized in Table~\ref{evolution}.  At the end of the Wolf-Rayet phase the stars explode imparting 10\,M$_{\odot}$ of material and 10$^{51}$\,ergs of thermal energy into the environment.  The lifespans of the stars are designed in such a way so that there are three distinct supernova explosions over the course of the simulation.

The simulation utilizes a temperature dependent average particle mass, and molecular, atomic and ionized phases are tracked separately.  The net heating/cooling rate per unit volume is parameterized as $\dot{e} = n\Gamma - n^2\Lambda$, where $\rm{n\,=\,\rho/m_H}$.  $\Gamma$ is the heating coefficient and is set at a constant value of $\Gamma = 10^{-26}\rm\,ergs\,s^{-1}$.  $\Lambda$ is the cooling coefficient which is assumed to depend only on temperature.  Cooling at low temperatures (T\,$\lesssim 10^4$\,K) is then adjusted to provide three thermally stable phases which correspond to the molecular (T\,$\sim$ 10\,K), atomic (T\,$\sim$ 150\,K) and ionized (T\,$\sim$ 8500\,K) phases.  This cooling curve and the phase diagram are shown in \citet{Pittard11}. Photoevaporation is not treated in these simulations.  However, as the photoevaporation time for a clump of these characteristics is comparable to the lifetime from ablation this should not significantly affect the results. 

The simulations in Paper I showed that the inhomogeneous structure of the natal GMC cloud surrounding the cluster had an important effect on the initial expansion of the cluster wind, which cut channels through the low density material to escape the clump.  The regions of high density within the initial clump proved to be surprisingly resistant to ablation from the cluster wind, and at later times the shockwaves of the SNRs.

\begin{table}
\centering
\caption{Wind properties of the three stars in the cluster as they evolve.}
\begin{tabular}{c@{~}c@{~}c@{~}c@{~}c@{~}c@{~}}
\hline
Stellar & \multicolumn{5}{c}{MS stage} \\
Mass    & $\dot{M}$ & v$_{\infty}$ & Duration & Mtm & Energy \\
($\rm\,M_{\odot}$) & ($\rm M_{\odot}\,\rm yr^{-1}$) & (km s$^{-1}$) & (Myr) & (kg m s$^{-1}$) & (ergs) \\
\hline
\hline
35 & 5.0$\times 10^{-7}$ & 2000 & 4.0 & 8.0$\times 10^{36}$ & 8.0$\times 10^{49}$ \\
32 & 2.5$\times 10^{-7}$ & 2000 & 4.5 & 4.5$\times 10^{36}$ & 4.5$\times 10^{49}$ \\
28 & 1.5$\times 10^{-7}$ & 2000 & 5.0 & 3.0$\times 10^{36}$ & 3.0$\times 10^{49}$ \\
\end{tabular}
\begin{tabular}{c@{~}c@{~}c@{~}c@{~}c@{~}c@{~}}
\hline
Stellar & \multicolumn{5}{c}{RSG stage} \\
Mass    & $\dot{M}$ & v$_{\infty}$ & Duration & Mtm & Energy \\
($\rm\,M_{\odot}$) & ($\rm M_{\odot}\,\rm yr^{-1}$) & (km s$^{-1}$) & (Myr) & (kg m s$^{-1}$) & (ergs) \\
\hline
\hline
35,32,28 & 1.0$\times 10^{-4}$ & 50 & 0.1 & 1.0$\times 10^{36}$ & 2.5$\times 10^{47}$ \\
\end{tabular}
\begin{tabular}{c@{~}c@{~}c@{~}c@{~}c@{~}c@{~}}
\hline
Stellar & \multicolumn{5}{c}{WR stage} \\
Mass    & $\dot{M}$ & v$_{\infty}$ & Duration & Mtm & Energy \\
($\rm\,M_{\odot}$) & ($\rm M_{\odot}\,\rm yr^{-1}$) & (km s$^{-1}$) & (Myr) & (kg m s$^{-1}$) & (ergs) \\
\hline
\hline
35,32,28 & 2.0$\times 10^{-5}$ & 2000 & 0.3 & 2.4$\times 10^{37}$ & 2.4$\times 10^{50}$ \\
\hline
\label{evolution} 
\end{tabular}
\end{table}

\subsection{Modelling the X-ray Emission and Absorption}
\label{sec:xray_modelling}
To calculate the X-ray emission the results from the hydrodynamical model are read into a radiative transfer ray-tracing code, and the appropriate emission and absorption coefficients are calculated for each cell using the temperature and density values.  A synthetic image on the plane of the sky is then generated by solving the radiative transfer equation along suitable lines of sight through the grid.  Solar abundances and collisional ionization equilibrium are assumed throughout this work.  The X-ray emissivity is calculated using the mekal emission code \citep[][and references therein]{Mewe95}.  The emissivity is stored in look-up tables containing 200 logarithmic energy bins between 0.1 and 10\,keV, and 91 logarithmic temperature bins between 10$^4$ and 10$^9$\,K.  Line emission dominates the emissivity at temperatures below 10$^7$\,K, with thermal bremsstrahlung dominating at higher temperatures.  The present calculations also have an interstellar absorption column (N$_{\rm H}\,=\,10^{21}$\,cm$^{-2}$) added to them, and each model is assumed to be at a distance of 1\,kpc from an observer.

The energy bins are split into three energy bands which represent the soft, medium and hard X-ray components of the spectra.  The soft X-ray regime runs from 0.1--0.5\,keV, the medium runs from 0.5--2.5\,keV and the hard X-rays run from 2.5--10.0\,keV.  These bands will be referred to as BB1, BB2 and BB3 respectively throughout this paper.

It should be noted that the individual stars are not resolved in the hydrodynamic simulations in Paper I, and therefore there is no contribution to the X-ray emission from the cluster wind interacting with any natal material close to the cluster \citep{Parkin10} or from intracluster wind-wind interactions \citep{Canto00,Pittard10b}.

\section{Results}
\label{sec:xray_results}
\subsection{The Main Sequence Phase}
\begin{figure}
\centering
\includegraphics[height=0.25\textwidth,width=0.49\textwidth]{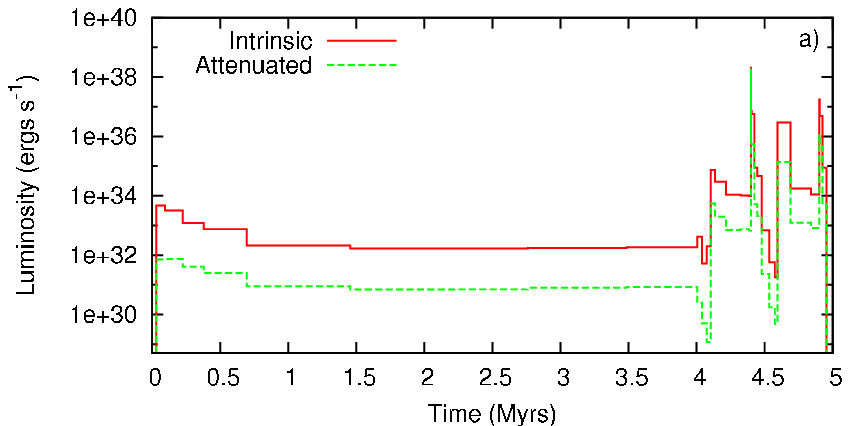}
\includegraphics[height=0.25\textwidth,width=0.49\textwidth]{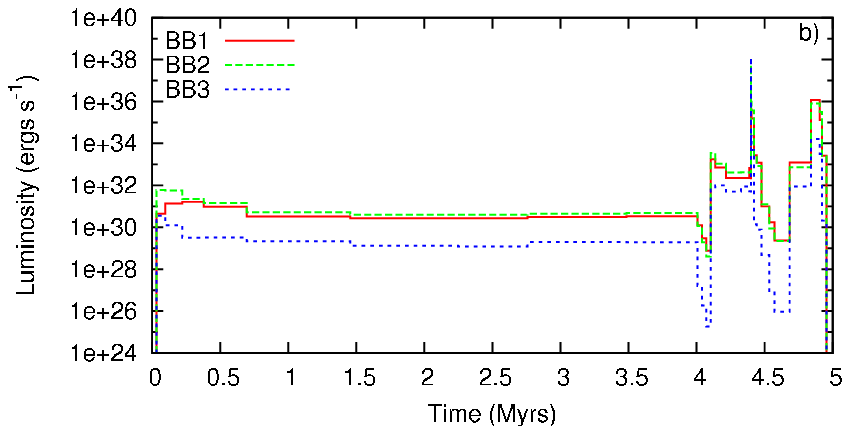}
\caption[The X-ray lightcurve for the cluster over the course of the simulation.]{The X-ray lightcurve for the cluster over the course of the simulation.  a) Shows the total intrinsic luminosity produced by the cluster (solid red line) compared with the observable luminosity after attenuation (dotted green line).  b) Shows the attenuated luminosity in all three energy bands defined in Sec~\ref{sec:xray_modelling}.  The solid red line shows the soft X-rays in BB1, the green dashed line shows the medium X-rays in BB2 and the blue dotted line shows the hard X-rays in BB3. \label{lightcurve}}
\end{figure}

The X-ray lightcurve for the cluster throughout the simulation is shown in the top panel of Fig.~\ref{lightcurve}.  The initial observable luminosity of the cluster is L$_X \sim$\,7$\times 10^{31}\rm\,ergs\,s^{-1}$.  Over the next 0.7\,Myrs this luminosity decreases by a factor of 10 to approximately L$_X\sim$\,9$\times 10^{30}\rm\,ergs\,s^{-1}$, at which point it remains fairly constant for the duration of the MS of the cluster.  Initially the X-ray luminosity is high as the cluster wind blown bubble is confined within the GMC clump.  However, as the wind blows out of the low density regions of the clump, hot gas escapes from the centre, as described by the ``leaky bubble'' model of \citet{Harper-Clark09} and Paper I.  The reduced pressure within the bubble caused by this leakage results in a lower X-ray luminosity. 

\begin{figure*}
\centering
\includegraphics[height=0.29\textwidth,width=0.29\textwidth]{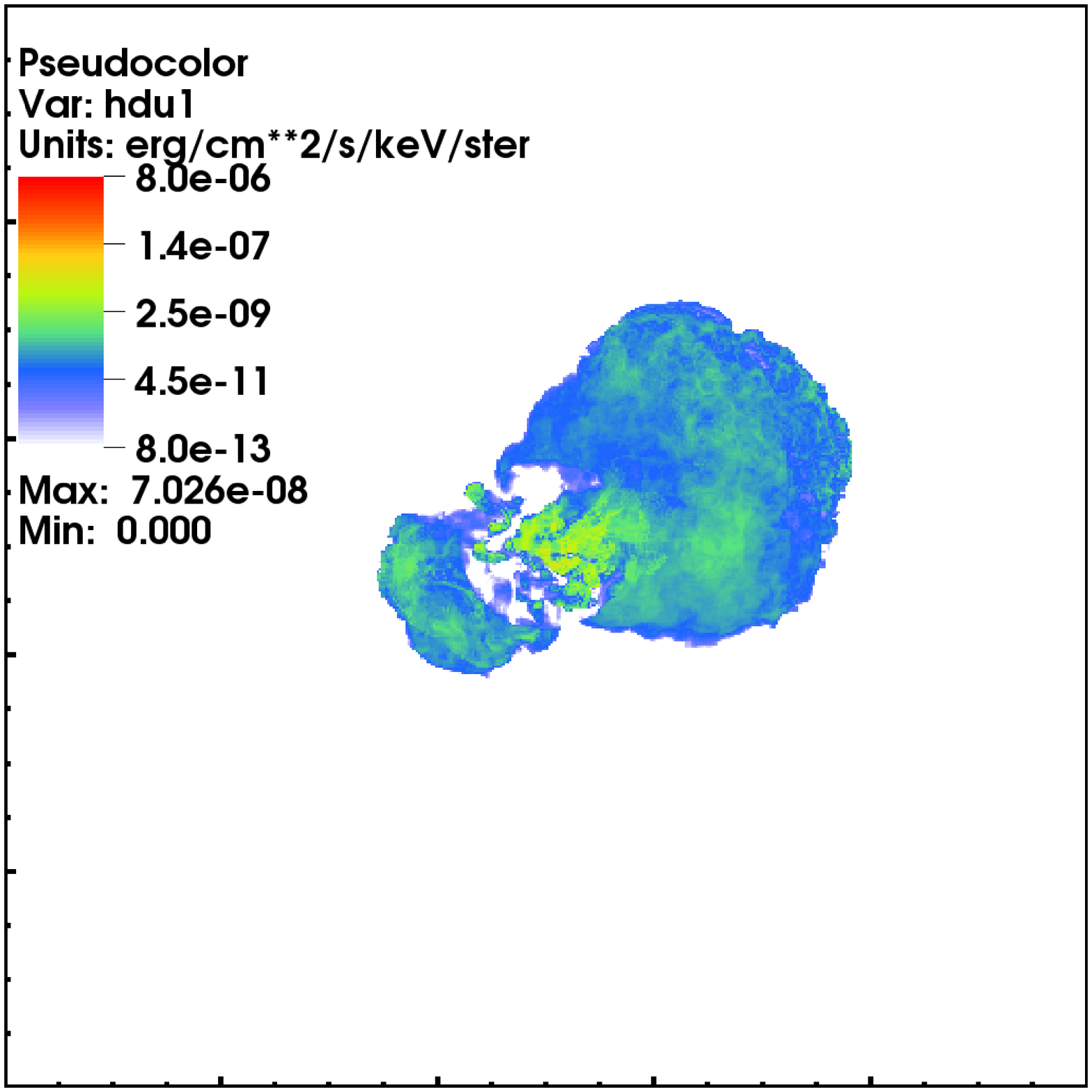}
\includegraphics[height=0.29\textwidth,width=0.29\textwidth]{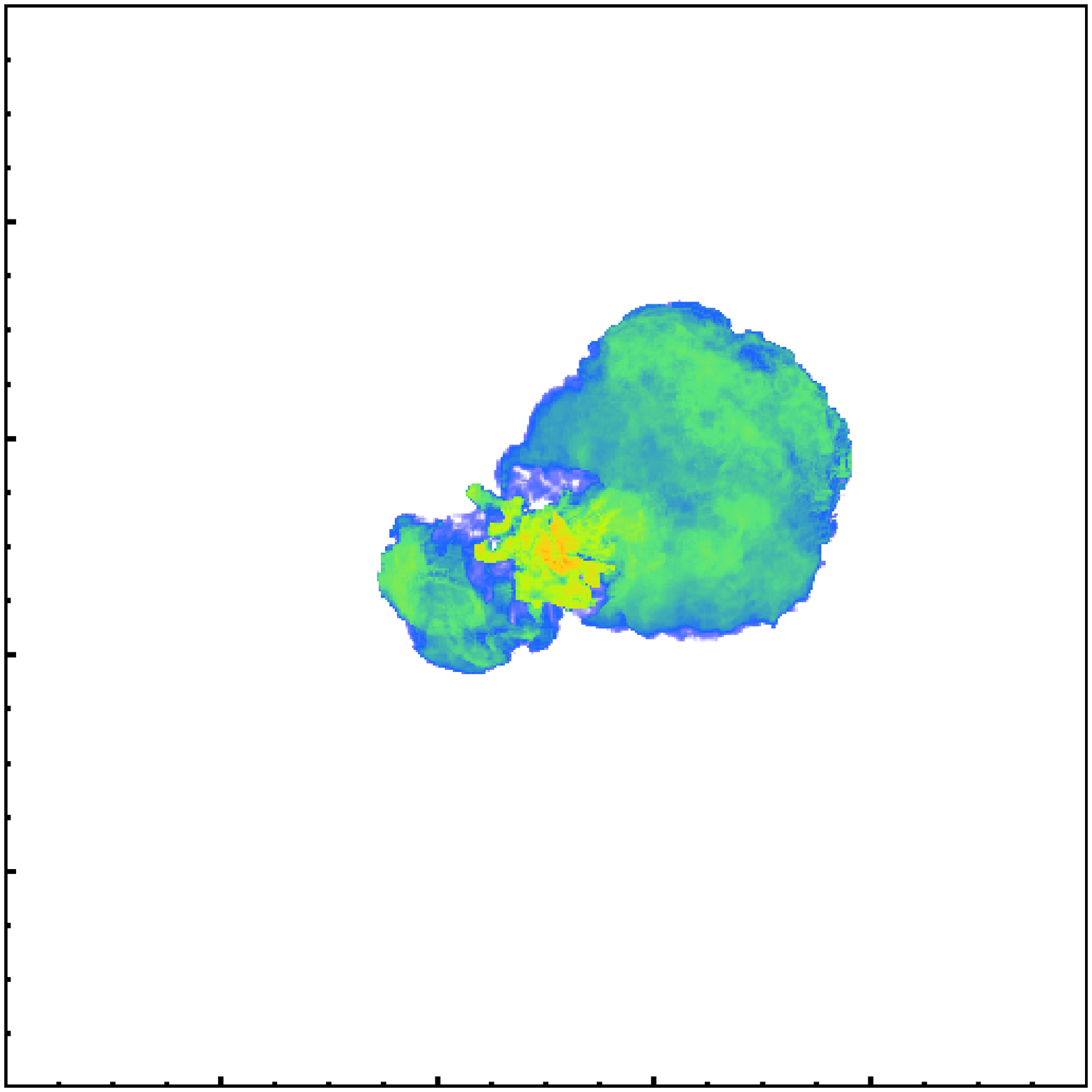}
\includegraphics[height=0.29\textwidth,width=0.29\textwidth]{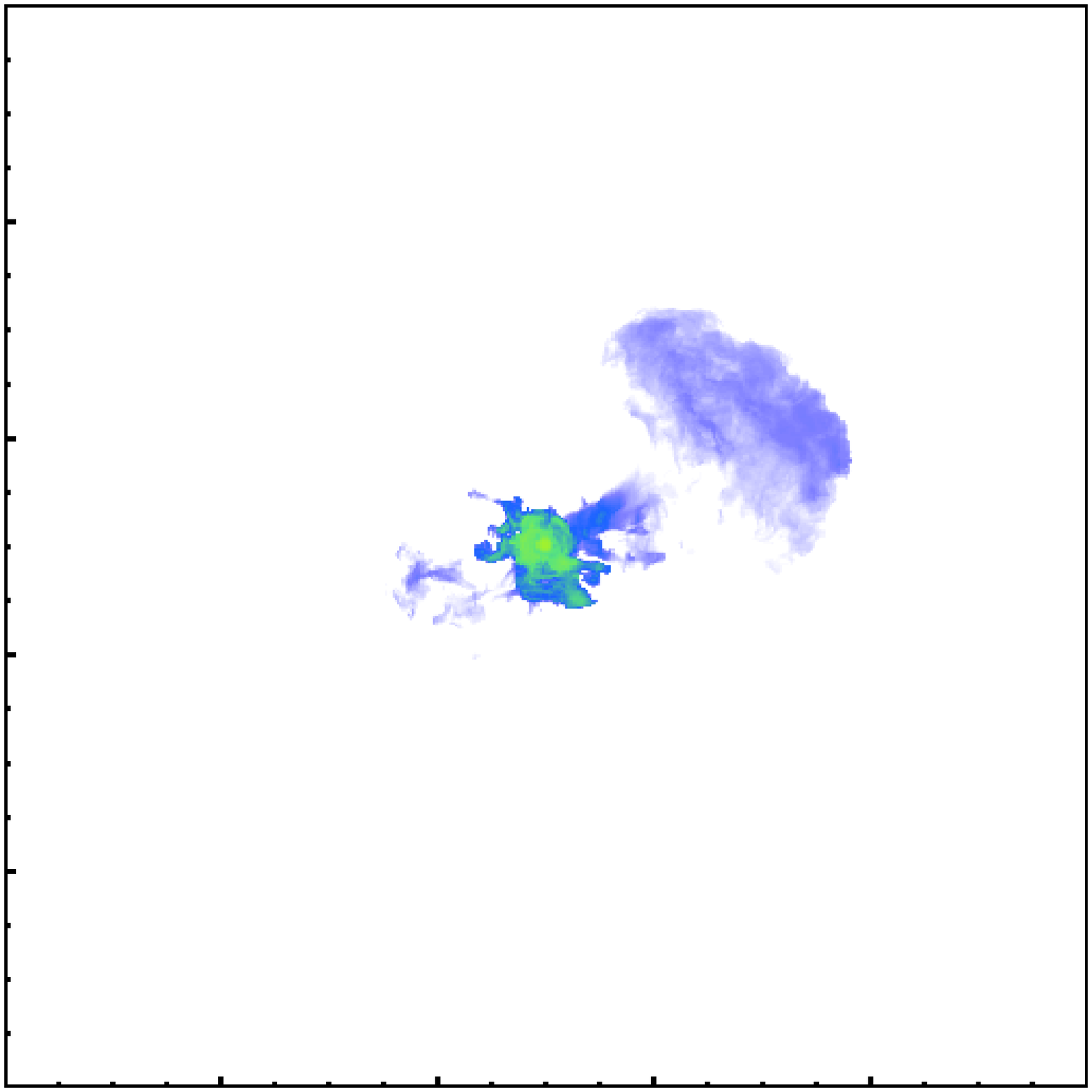}
\caption[X-ray emission for the cluster at time t\,=\,0.13\,Myrs.]{X-ray emission for the cluster at time t\,=\,0.13\,Myrs.  Each panel has sides of 500 pixels and length 55.4\,pc. [Left] shows soft X-rays 0.1--0.5 keV, [Middle] shows medium X-rays at 0.5--2.5 keV and [Right] shows hard X-rays at 2.5--10.0 keV.  The stellar cluster is at the centre of each panel. \label{xray_04}}
\end{figure*}

Fig.~\ref{xray_04} shows simulated X-ray images of the cluster at time t\,=\,0.13\,Myrs, where extended bubbles to either side indicate that some of the hot wind material is leaking from the GMC clump.  However, it is clear that there is still partial confinement by the inhomogeneous GMC clump since the images are not spherically symmetric.  At this time all three stars are on the MS (see Table~\ref{evolution} for the stellar properties).  The left and middle panels show images of the soft and medium energy X-rays produced at this time, whilst the right panel shows the hard X-rays.  The emission is brightest at the centre in all three images, but particularly so in the medium and hard images.  At this early time there is strong absorption of the soft X-rays within the GMC clump, as revealed by the low surface brightness of regions which are more clearly emitting at higher energies (compare the left and middle panels).  This behaviour is not so prominent in the medium energy X-ray image, although there is some absorption occuring.

The most striking feature in the images is the extended emission to the top right of the cluster, which results from the hot gas that has already broken out of the clump in this direction.  It is also interesting to observe that the hot fluid adiabatically cools as it accelerates to supersonic speeds through the `nozzles' from which it leaves the confining clump.  This is visible as a reduction in the X-ray surface brightness in the hard band.  The surface brightness of this gas increases at larger distances from the clump as it passes through a termination shock.  At this point it runs up against previously shocked gas which is inflating the bubble and sweeping up a shell of the ambient medium which surrounds the GMC clump.  The hardest X-rays are produced by the hot gas in the cluster centre, where the cluster wind is partially confined.  The bottom panel of Fig.~\ref{lightcurve} shows that the luminosities in the BB1 and BB2 energy bands are almost equal during the period when all three stars are on the MS.  However, at t$\,<\,0.25\rm\,Myrs$ the luminositiy in the BB2 energy band is dominant due to the greater attenuation of the lower energy X-rays by the remnant GMC clump.  The hard X-ray luminosity in the BB3 energy band is about an order of magnitude lower than the luminosities in the soft and medium energy bands throughout the MS-dominated phase of the cluster evolution.

At t\,=\,0.13\,Myrs, approximately 90$\%$ of the total (0.1-10\,keV) intrinsic X-ray luminosity originates from the inner 4\,pc radius of the simulation, which is the original radius of the GMC clump containing the cluster.  This is not unexpected as the cluster wind is young and hot plasma vents out of only a few open channels at this time.  Since the GMC clump is mostly intact, significant attenuation of low energy X-rays occurs within the clump radius.  This is reflected by the fact that only $\sim$ 2/3 of the total attenuated luminosity originates from the inner 4\,pc radius, indicating that the dense material within that radius has absorbed a substantial amount of the intrinsic emission.

The low density regions of the clump are rapidly blown out by the cluster wind but afterwards the remaining high density regions are much longer-lived.  It should be noted that in simulations where there is hot, low density material in contact with cold, high density material minor heating of the cold dense gas may occur, and will be a function of the simulation resolution \citep{Parkin10}.  The effective covering fraction of the densest regions, the gross properties of the cluster wind and the X-ray luminosity evolve only slowly.  However, the characteristic size of the emitting region continues to increase.  Although much of the hot gas leaves the grid through the outflow boundaries, we can nevertheless examine how the emission from hot gas on the grid evolves with time.  At t\,=\,0.44\,Myrs approximately 50$\%$ of both the intrinsic and attenuated luminosity originates from within the original clump radius.  By t\,=\,1.96\,Myrs this value has decreased to 20$\%$ for both luminosities and by t\,=\,2.53\,Myrs, approximately midway through the MS, only 12$\%$ of the luminosity originating on the numerical grid comes from the central 4\,pc radius.  This decline is driven by two factors.  First, the cluster wind increasingly clears out dense molecular gas within the orignal GMC clump as time goes on.  The dense gas is both ablated into the hot gas streaming past and also pushed away.  The bowshocks which form around the nearest dense clouds, and which merge to produce the reverse/termination shock of the cluster wind, thus form at greater and greater distances.  Hence there is simply less hot gas in this central region as time increases.  Secondly, the ablated material is entrained into the outflows away from the GMC clump.  Indeed, mass entrainment/loading factors may exceed $\sim$100 (see Section 3.4 in Paper I).  This entrainment increases the density of the flow and its emissivity, while also creating slow moving obstacles which faster moving parts of the flow shock against.  The flow outside of the GMC clump thus contains a multitude of shocks, and a wide range of densities and temperatures.

\begin{figure*}
\centering
\includegraphics[height=0.29\textwidth,width=0.29\textwidth]{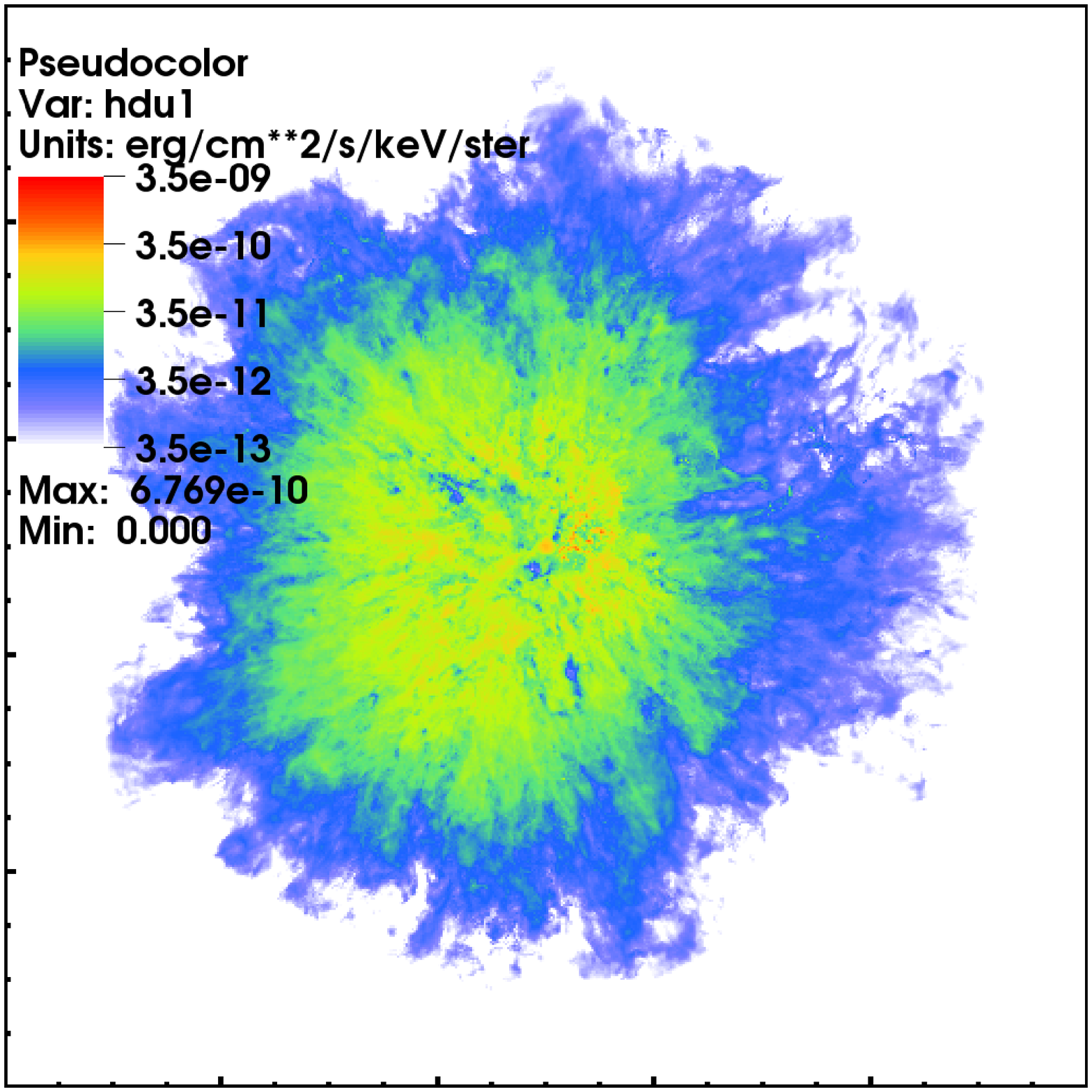}
\includegraphics[height=0.29\textwidth,width=0.29\textwidth]{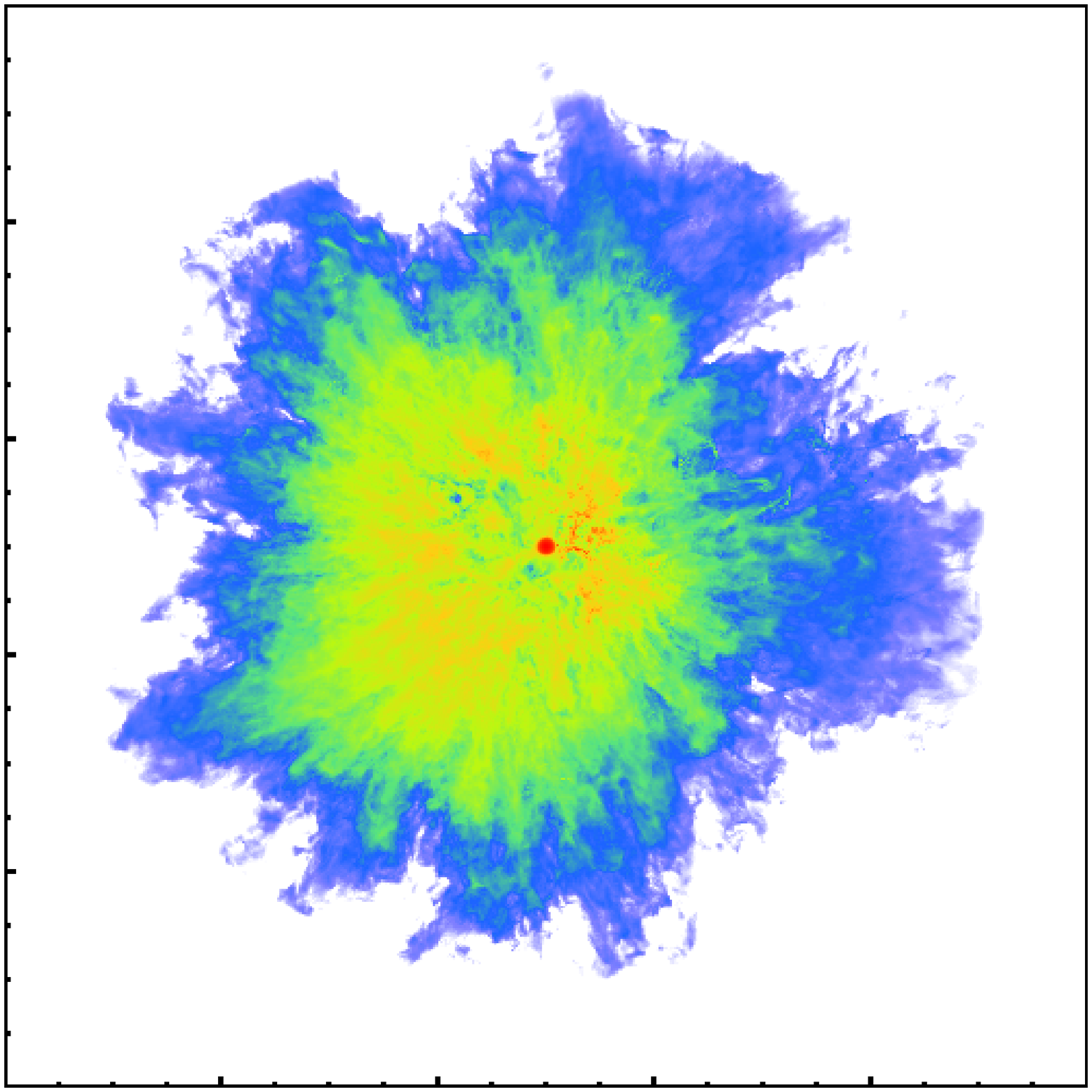}
\includegraphics[height=0.29\textwidth,width=0.29\textwidth]{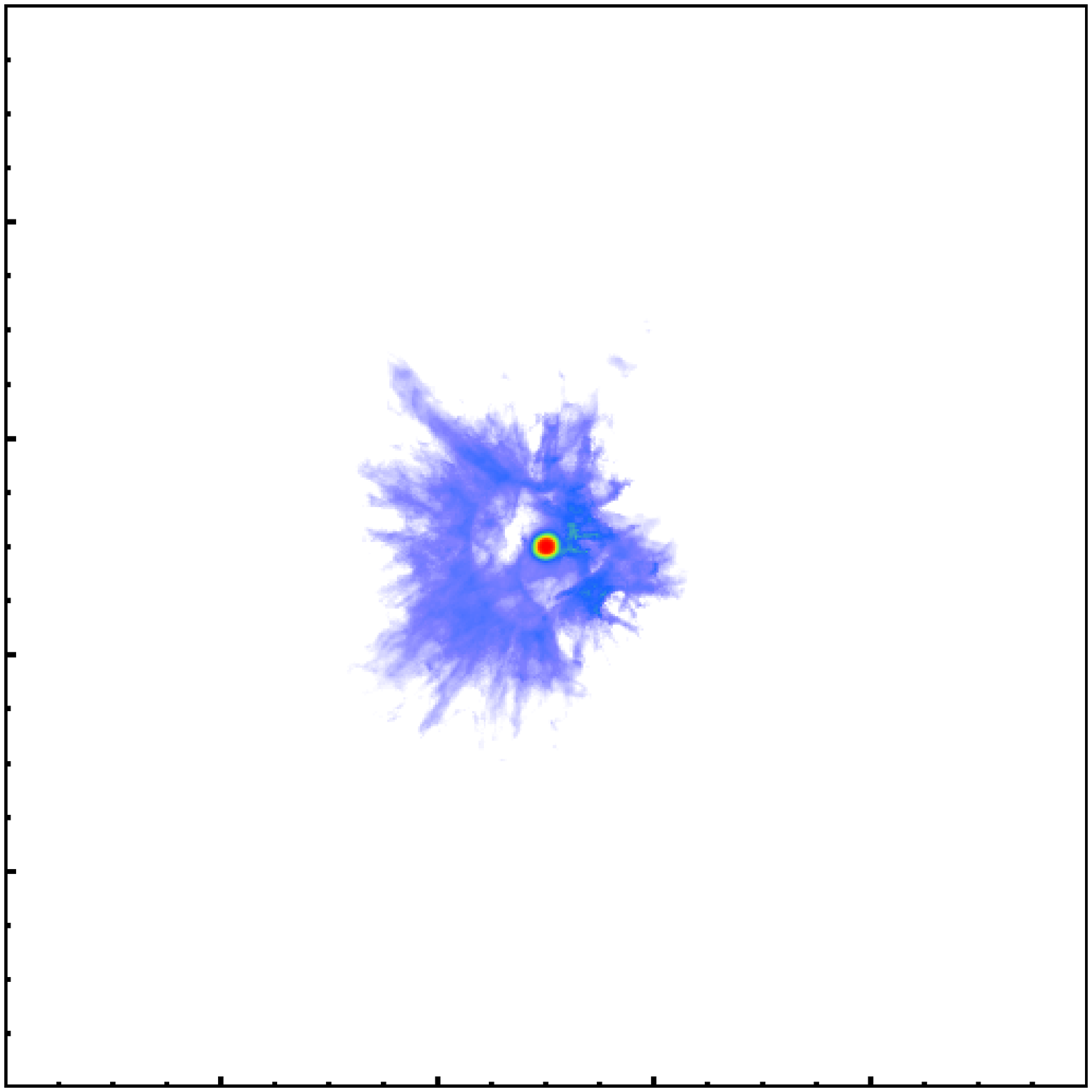}
\caption[X-ray emission for the cluster at time t\,=\,2.53\,Myrs.]{X-ray emission for the cluster at time t\,=\,2.53\,Myrs.  Each panel has sides of 500 pixels and length 55.4\,pc. [Left] shows soft X-rays 0.1--0.5 keV, [Middle] shows medium X-rays at 0.5--2.5 keV and [Right] shows hard X-rays at 2.5--10.0 keV. \label{xray_80}}
\end{figure*}

X-ray images of the cluster at t\,=\,2.53\,Myrs are shown in Fig.~\ref{xray_80}.  There is considerable diffuse emission in the soft and medium X-ray bands (left and middle panels of Fig.~\ref{xray_80}).  In contrast, the spatial extent of the hard X-rays is much smaller, and these instead primarily trace the stellar cluster and the hot, shocked gas immediately downstream of the reverse shock of the cluster wind.  The gas responsible for this emission reduces in temperature as colder material from the remains of the GMC clump mixes in with it, which limits the extent of the emission in this image.  The extent of the diffuse emission reaches well beyond the original cluster radius, with more than 88$\%$ of the overall luminosity originating from outside that radius. Although the images for the soft and medium regimes are similar in structure, with strong emission in the centre and a filamentary diffuse structure towards the edges, there is a higher intensity in the BB2 image, and the stellar cluster is clearly discernable at the centre of the clump.

\subsection{ISM Absorption Effects}
\begin{figure}
\centering
\includegraphics[height=0.25\textwidth,width=0.49\textwidth]{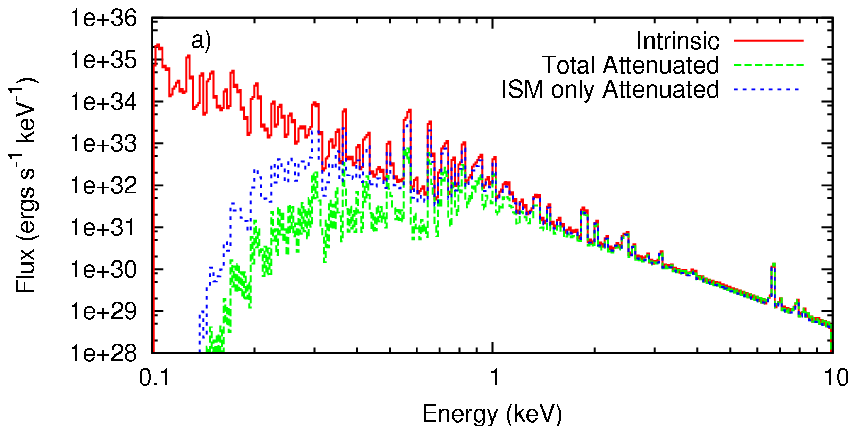}
\includegraphics[height=0.25\textwidth,width=0.49\textwidth]{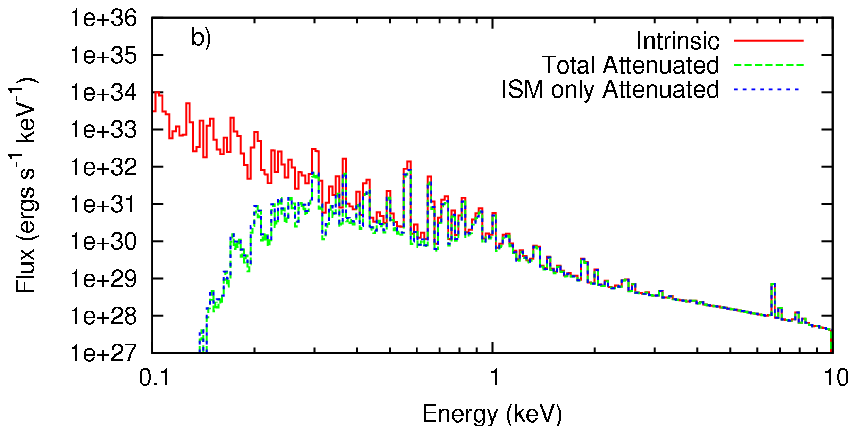}
\caption[X-ray spectra of the cluster at two times during the MS dominated phase.]{X-ray spectra of the cluster at two times during the MS dominated phase.  The red solid line shows the total intrinsic emission produced by the cluster.  The green dashed line shows the total observable emission after all attenuation effects are considered, whilst the blue dotted line shows the effect of just the ISM absorption on the emission.  a) Shows the spectra at t\,=\,0.06\,Myrs, when there is a notable difference between the ISM only and total attenuated emission.  b) Shows the spectra at t\,=\,2.53\,Myrs when absorption by dense material from the GMC clump has little effect on the overall attenuated emission. \label{ISM_absorption}}.
\end{figure}

The attenuated X-ray luminosity is dependent on both the ISM column density and the density and size of the GMC clump in which the cluster forms.  However, as the molecular material in the clump is ablated by the winds it will have less of an effect on the observable luminosity of the cluster.  Fig.~\ref{ISM_absorption} gives an indication of the degree of attenuation caused by the ISM and by dense clump material.  The red solid line shows the intrinsic X-ray spectrum from the cluster, with the green dashed line showing the total attenuated spectrum taking all absorption effects into account.  As discussed previously, the vast majority of the absorption occurs at soft X-ray energies, with very little occuring above E\,=\,1.0\,keV.  The blue dotted line shows the attenuation effects caused only by absorption from the ISM.  Close inspection of Fig.~\ref{ISM_absorption} reveals that the ISM absorption has little effect above 0.5\,keV, whereas the circumcluster absorption affects the spectrum up to energies around 1\,keV.  Therefore it is clear that these two distinct absorption components affect the spectrum in slightly different ways.  It is consistent with the hottest gas (and therefore the hardest emission) being buried more deeply within the GMC clump.  We note that the average column density from the centre of the cluster through the GMC clump to an observer at t\,=\,0.06\,Myrs is, at $\approx$\,3$\times10^{21}\rm\,cm^{-2}$ (see Fig.12 in Paper I), about 3 times the assumed ISM column.

At t\,=\,2.53\,Myrs (bottom panel of Fig.~\ref{ISM_absorption}) the two attenuated spectra are practically identical.  This is because the X-ray emitting gas is no longer confined by the dense absorbing material of the GMC clump as it is in the early stages of the bubble's expansion, but now suffuses through the entire volume of the simulation.  This is again consistent with the average column density from the centre of the cluster through the GMC clump to an observer at this time, which Fig.12 from Paper I shows to be about 10$^{19.6}\rm\,cm^{-2}$, or only about 4$\%$ of the assumed ISM column.

Changing the viewing angle to the cluster at late times leads to only very small differences ($\approx$ 5--10\,$\%$) in the attenuated luminosity, reinforcing the conclusion that the destruction of the clumpy environment results in very minimal column densities and thus attenuation by this gas.  Even at earlier times when the hot wind gas is still breaking out of the clump the difference in the attenuated luminosity as the viewing angle to the cluster changes is only around 15--20\,$\%$.  This likely reflects the relatively low initial column density of the clump.  Larger variations can be expected from models with higher initial column densities.

\subsection{RSG and WR Phases for the 35\,M$_{\odot}$ Star}
At t\,=\,4.0\,Myrs the most massive star evolves into a RSG.  Its mass loss rate increases, and its wind velocity decreases.  The averaged mass-loss rate and speed of the cluster wind then changes from 9$\times 10^{-7}\rm\,M_{\odot}\,yr^{-1}$ and 2000\,km\,s$^{-1}$ to $\approx\,10^{-4}\rm\,M_{\odot}\,yr^{-1}$ and 136\,km\,s$^{-1}$.  The cluster wind therefore becomes slow and dense.  The central cluster is no longer a source of hard X-rays, and there is no replenishment of the highest temperature gas in the surrounding environment as it flows away from the cluster through the remains of the porous GMC clump (see Fig.8 in Paper I).  Together these changes lead to a substantial reduction in the amount of hard X-rays being produced.  In fact, the BB3 luminosity decreases 4 orders of magnitude from L$_X \sim 2\,\times 10^{29}\rm\,ergs\,s^{-1}$ just before the evolutionary transition to L$_X \sim 2\,\times 10^{25}\rm\,ergs\,s^{-1}$ by the end of the RSG phase (see Fig.~\ref{lightcurve}).  In contrast, the intrinisic luminosity briefly increases following this transition, due to an increase in luminosity in the BB1 band.

Whilst the reason for this is not completely understood, it is possible that the sudden drop in pressure during the transition causes material stripped from the dense clouds to mix more rapidly with the hotter gas.  Overall however, because the soft X-rays suffer from attenuation from the dense RSG-enhanced wind material close to the centre of the cluster, the attenuated emission actually drops.  

The dense material deposited during the cluster wind's first RSG phase is subsequently cleared from the simulation volume once the most massive star further evolves to its WR phase at t\,=\,4.1\,Myrs.  The combined average speed of the cluster wind increases back to 2000\,km\,s$^{-1}$ while the combined mass-loss rate becomes 2.04$\times 10^{-5}\rm\,M_{\odot}\,yr^{-1}$.  The very high momentum that the cluster wind now has efficiently clears out the RSG dominated cluster wind filling the lower density channels and dramatically increases the ablation rate of the remaining dense clouds.  This causes increased emission in all three X-ray bands.  The intrinsic luminosity increases by over 2 dex above that reached in the MS phase, with an initial peak that then declines quickly to a steady value.  This phase is relatively short-lived, lasting only 0.3\,Myrs. 

\subsection{The First Supernova}
The most massive star explodes as a supernova at t\,=\,4.4\,Myrs, imparting 10\,M$_{\odot}$ of ejecta and 10$^{51}$\,ergs of thermal energy into the centre of the GMC clump.  At this point both of the other stars remain in their MS phases.

\begin{figure}
\centering
\includegraphics[height=0.25\textwidth,width=0.49\textwidth]{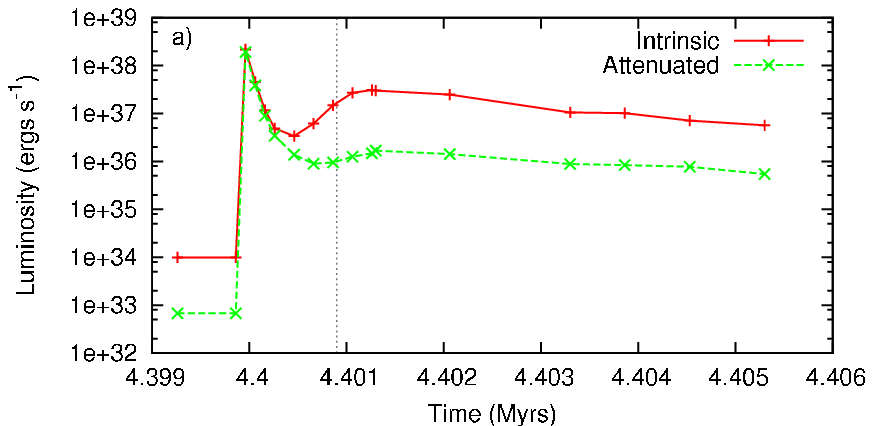}
\includegraphics[height=0.25\textwidth,width=0.49\textwidth]{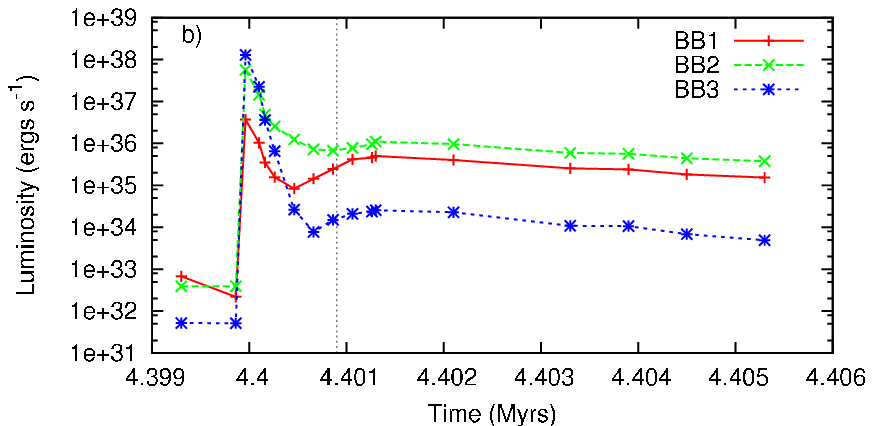}
\caption[The X-ray light curve for the cluster at the time of the first SN explosion.]{The X-ray light curve for the cluster at the time of the first SN explosion.  The black dashed line indicates 900 years after the explosion, at which point emission from interactions with the surrounding clump material is dominant in all three energy bands. a) Shows the total intrinsic luminosity produced by the cluster (solid red line) compared with the observable luminosity after attenuation (green dashed line).  b) Shows the attenuated luminosity in all three of the broadband energy bands.  The solid red line shows the soft X-rays in BB1, the green dashed line shows the medium energy X-rays in BB2 and the blue dotted line shows the hard X-rays in BB3. \label{snelightcurve}}
\end{figure}

The X-ray lightcurve immediately following the supernova explosion is shown in Fig.~\ref{snelightcurve}.  The SN ejecta is highly overpressured and rapidly expands into the surrounding medium.  Although this approach leads to the desired response on the surrounding medium, in actual SN explosions the ejecta rapidly cools through adiabatic expansion, and is considerably cooler than the simulated ejecta at comparable times.  Therefore the bright peak in the X-rays seen in Fig.~\ref{snelightcurve} immediately after the explosion should be ignored as it is an artifact of the utilized approach.  The X-ray luminosity of the hot ejecta drops rapidly from its peak as the ejecta starts to expand and its density decreases.  However, Fig.~\ref{snelightcurve} shows that the rate of decline of the X-ray luminosity decreases, and a minimum is reached after which the X-ray luminosity increases again.  This behaviour is caused by ejecta running into the remaining dense clouds near the cluster.  The kinetic energy that this ejecta has acquired at this time is then re-thermalized and subsequently radiates more strongly.  The dashed line 900\,yrs after the explosion indicates when the luminosity is dominated by the interactions of the ejectra with surroundng gas, and thus no longer affected by the explosion setup.

\begin{figure*}
\centering
\includegraphics[width=0.3\textwidth, height=0.3\textwidth]{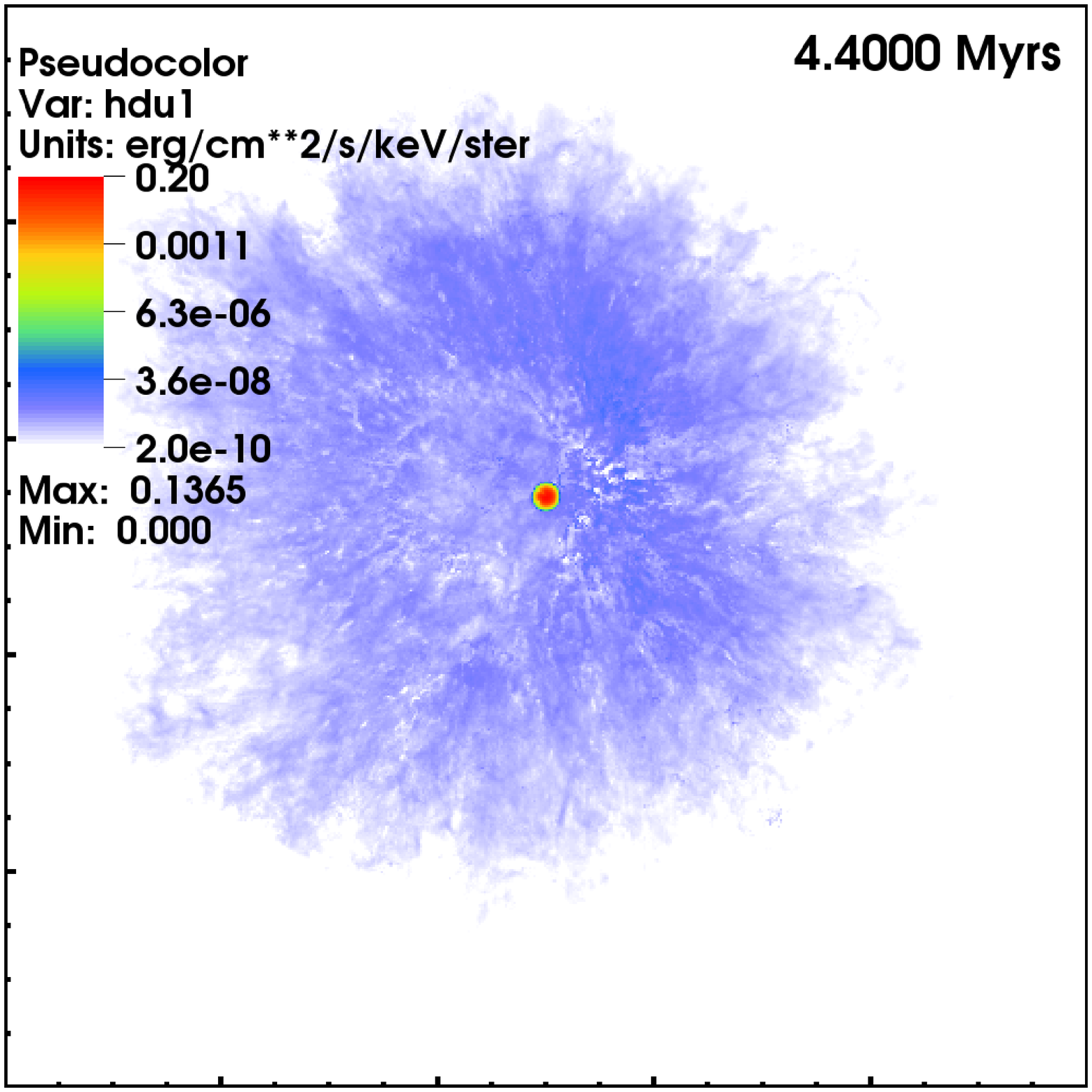}
\includegraphics[width=0.3\textwidth, height=0.3\textwidth]{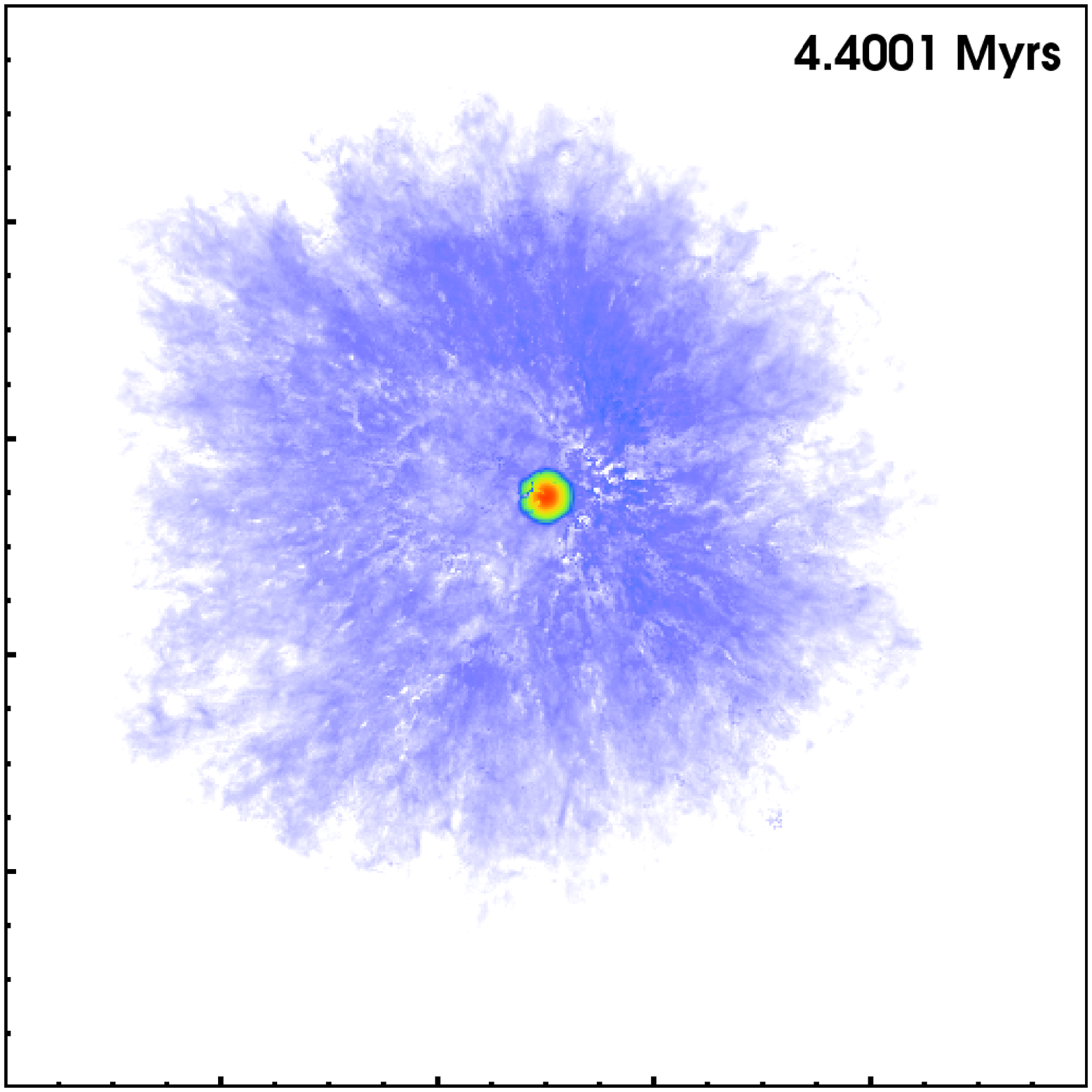}
\includegraphics[width=0.3\textwidth, height=0.3\textwidth]{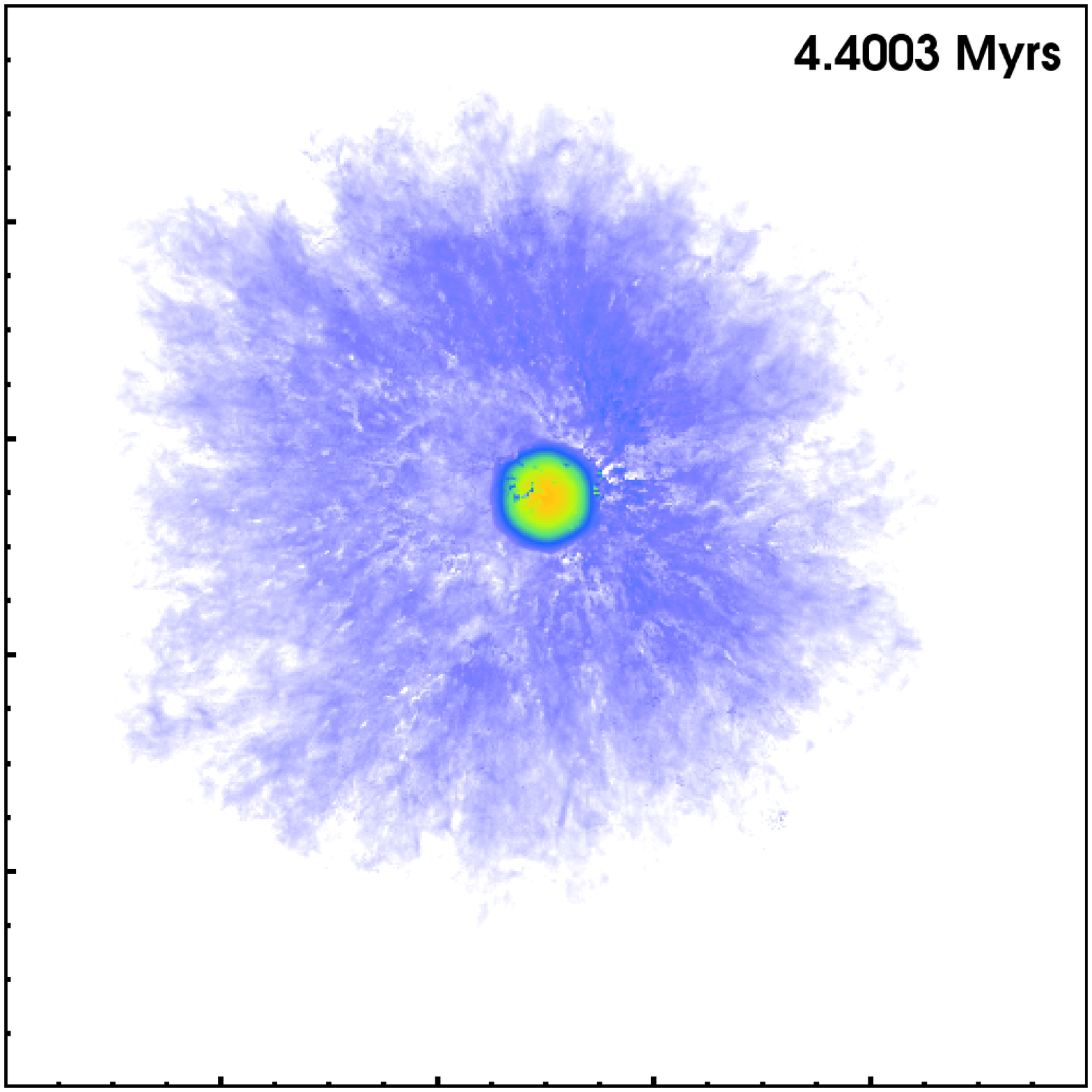}
\includegraphics[width=0.3\textwidth, height=0.3\textwidth]{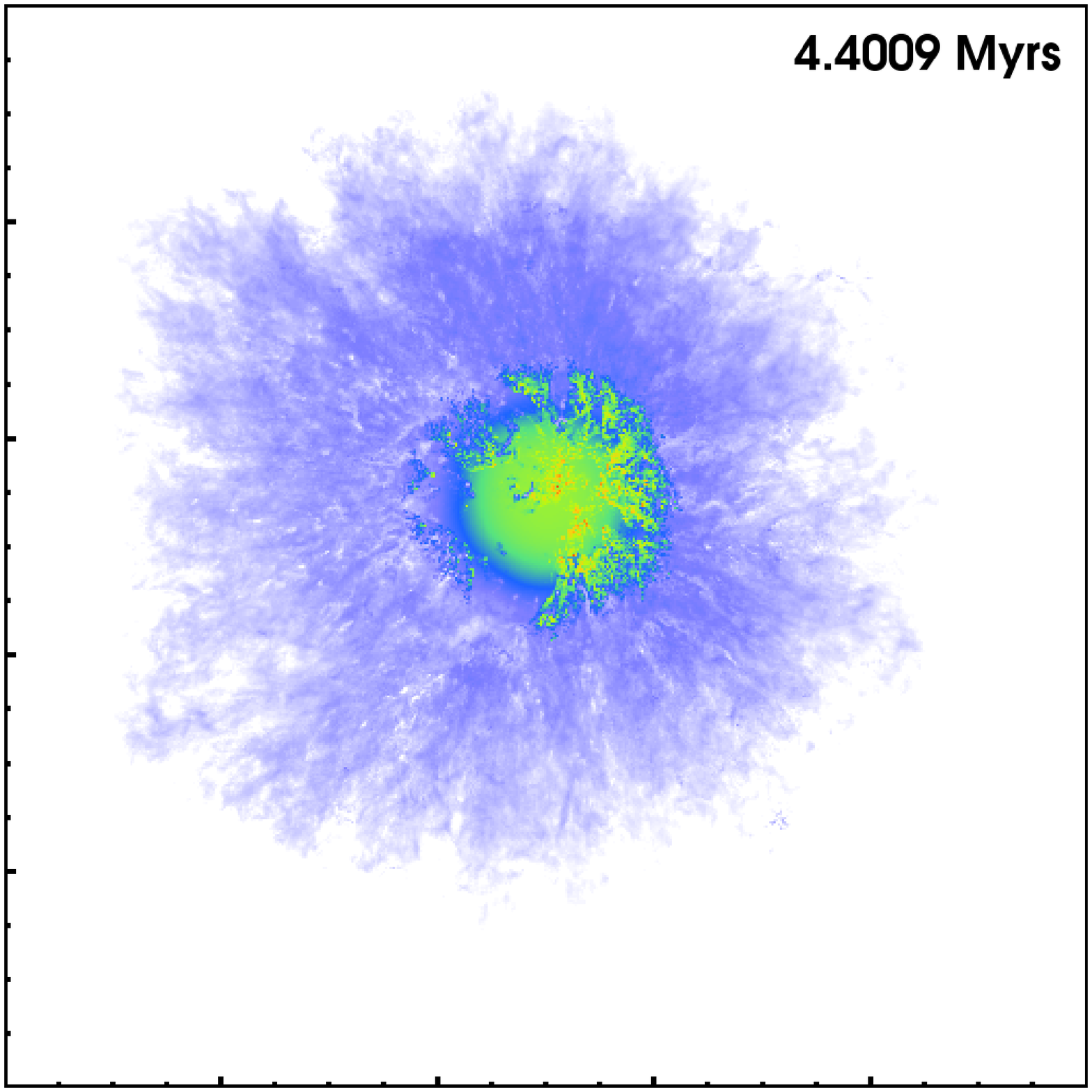}
\includegraphics[width=0.3\textwidth, height=0.3\textwidth]{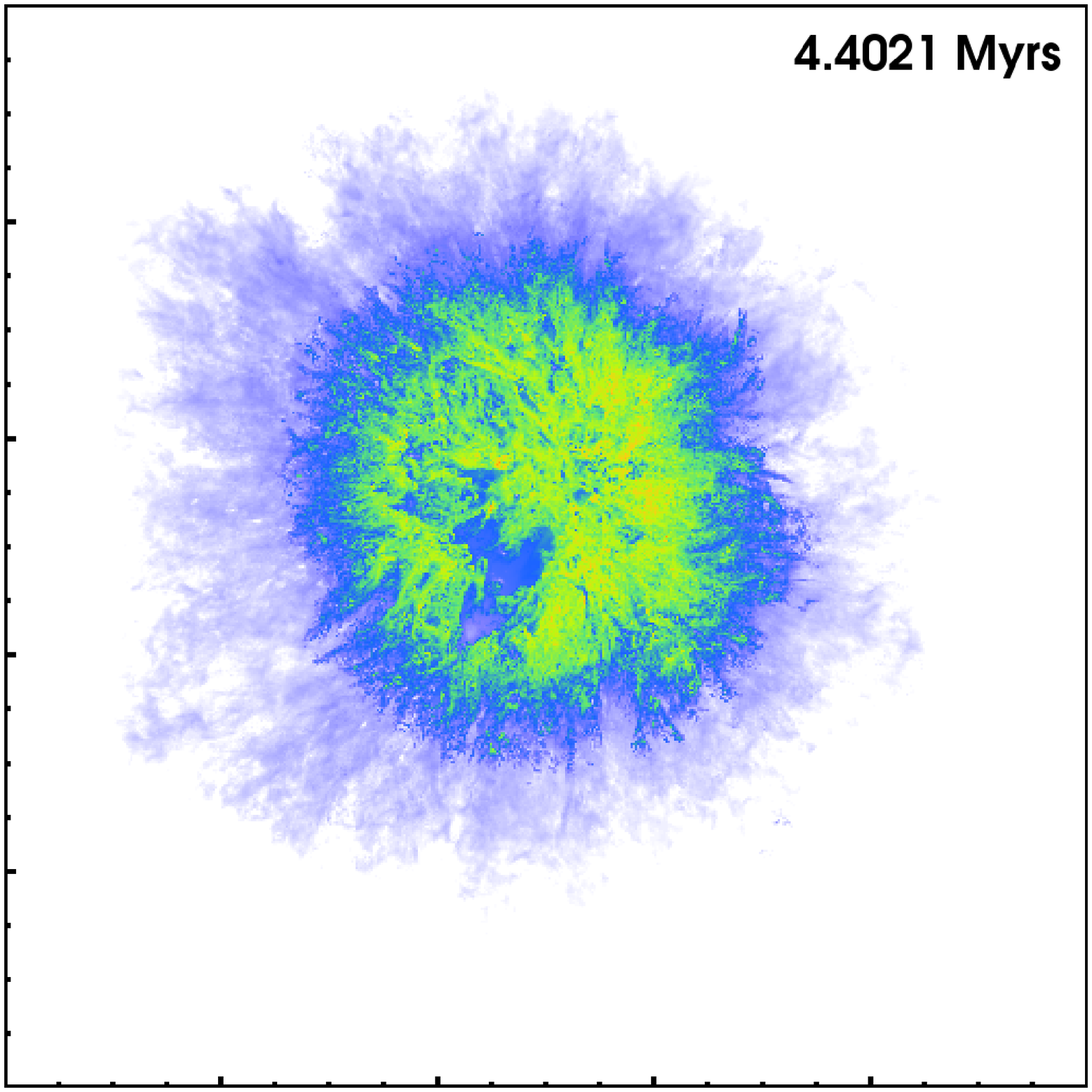}
\includegraphics[width=0.3\textwidth, height=0.3\textwidth]{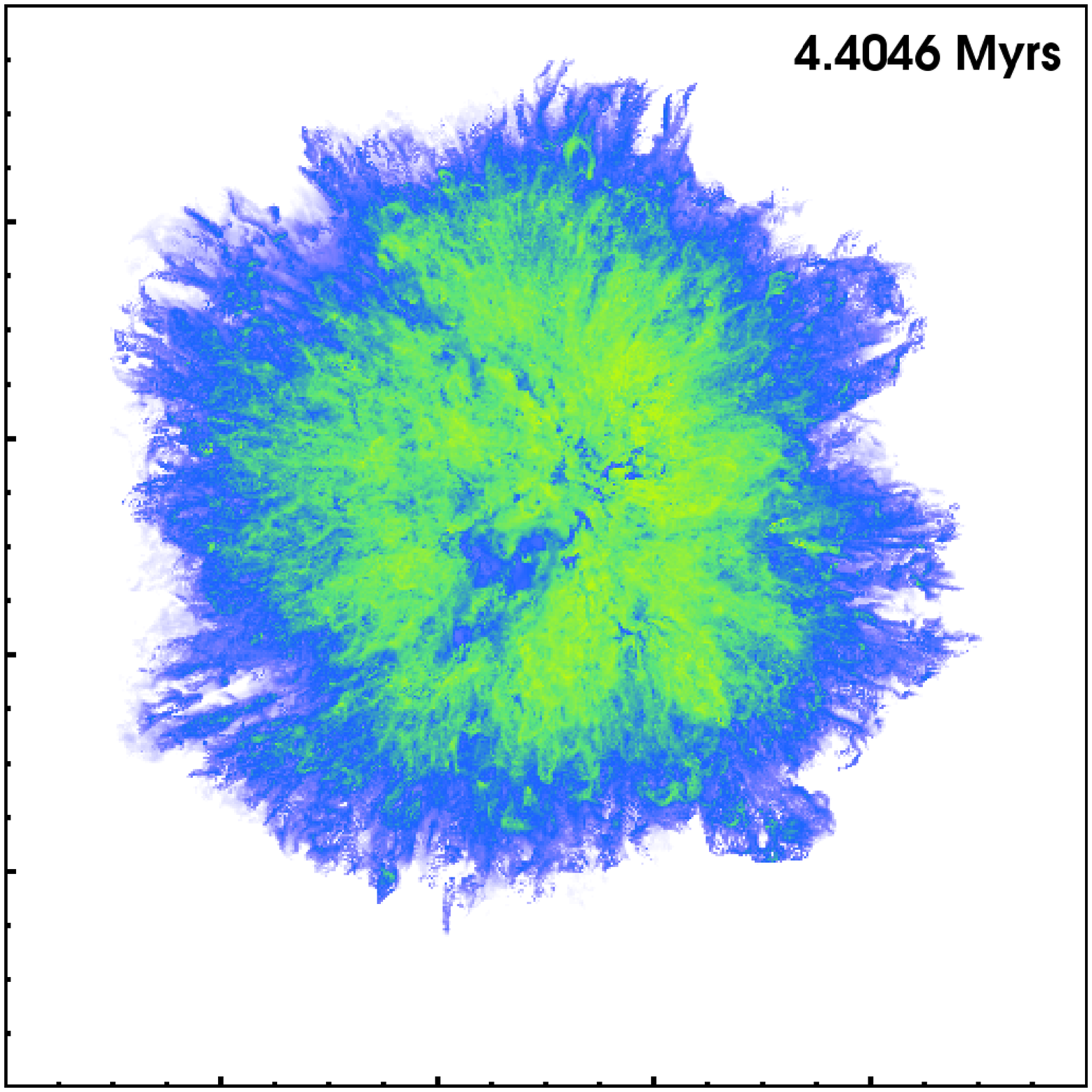}
\caption[Synthetic X-ray image in the BB1 (0.1--0.5\,keV) region during the first 4600\,yrs after the most massive star explodes.]{Synthetic X-ray image in the BB1 (0.1--0.5\,keV) energy band during the first 4600 years after the most massive star explodes.  The explosion occurs at t=4.4000\,Myrs (top left panel).  Absorption is visible from 100 years after the explosion.  Bow shock emission dominates from 900 years. \label{bb1_sne1}}
\end{figure*}

\begin{figure*}
\centering
\includegraphics[width=0.3\textwidth, height=0.3\textwidth]{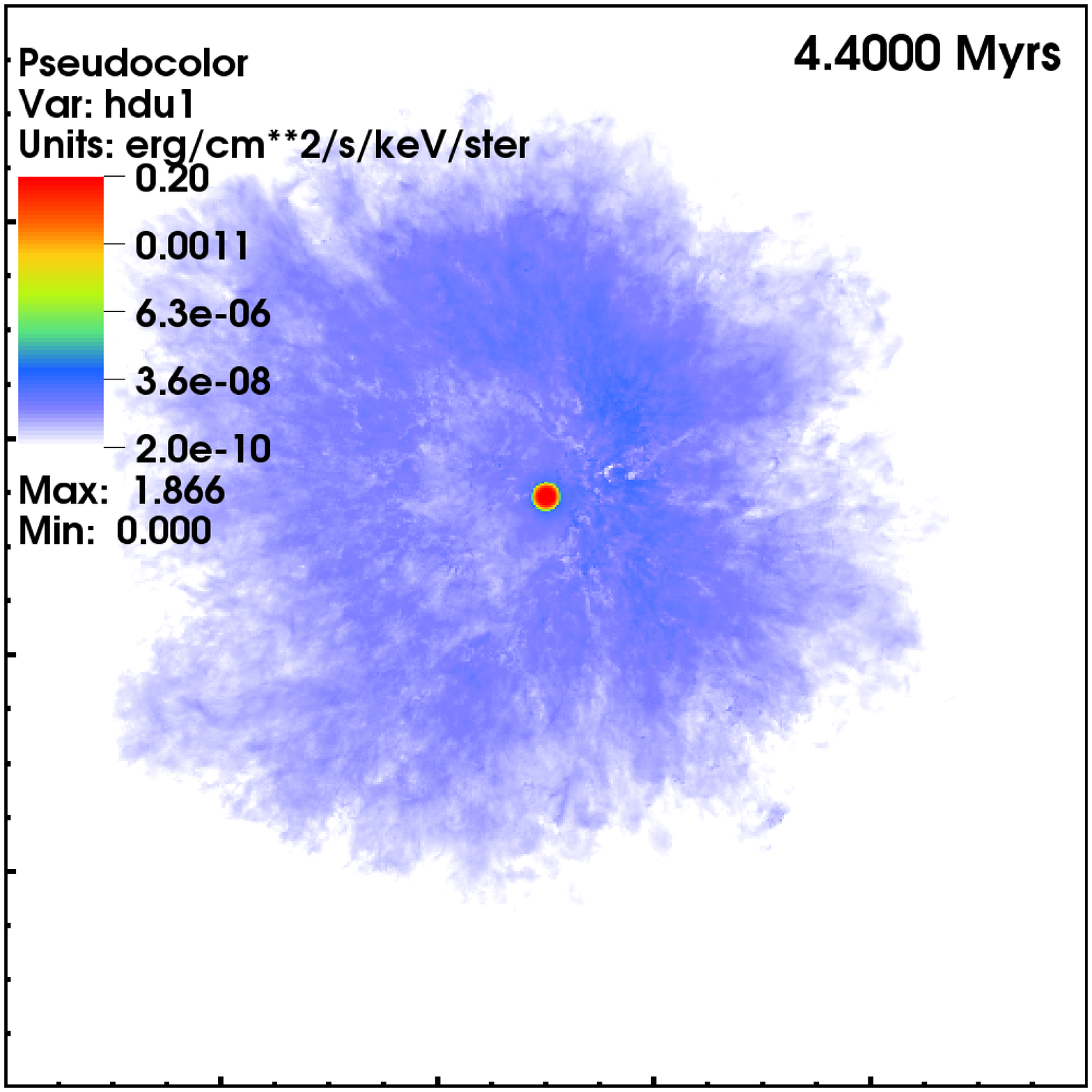}
\includegraphics[width=0.3\textwidth, height=0.3\textwidth]{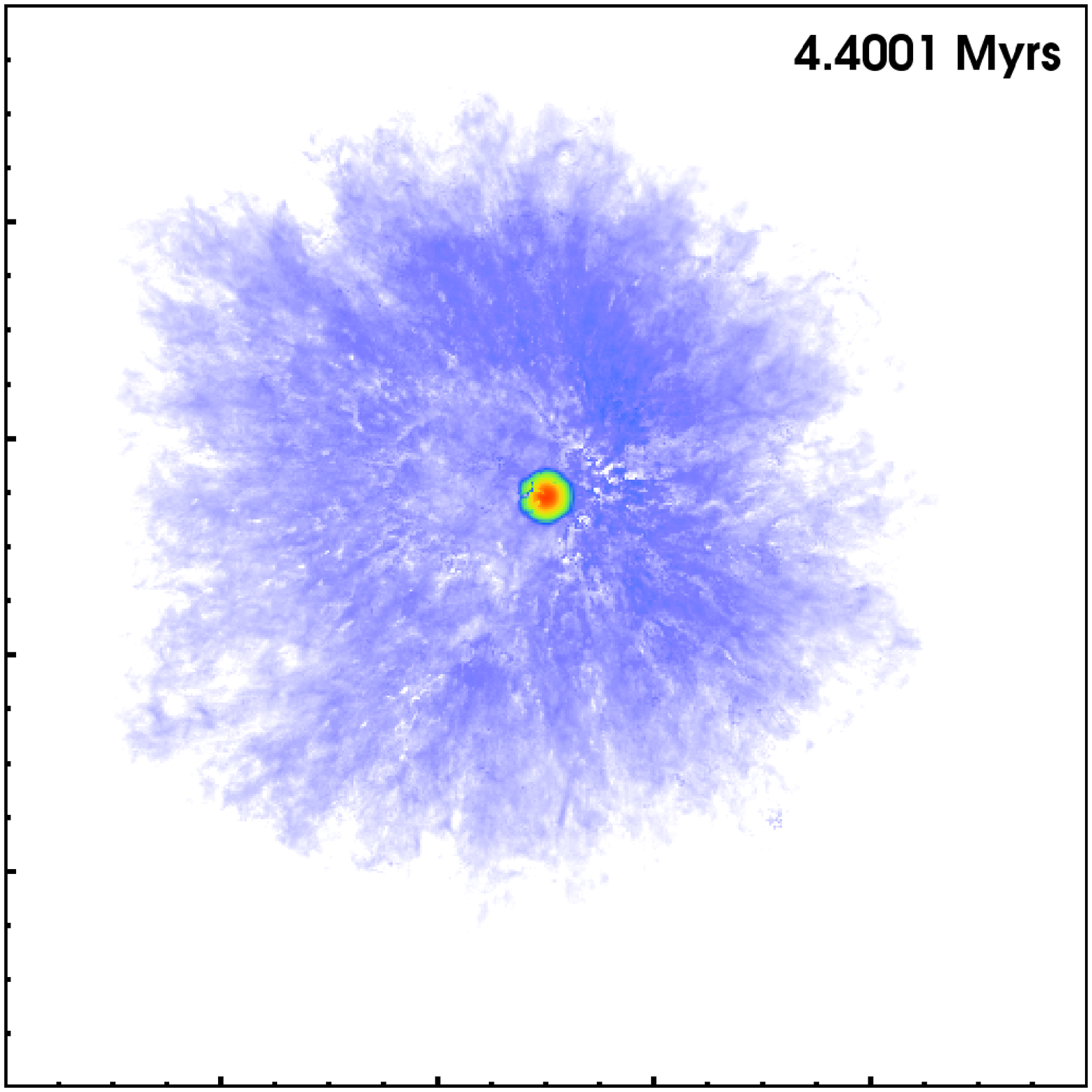}
\includegraphics[width=0.3\textwidth, height=0.3\textwidth]{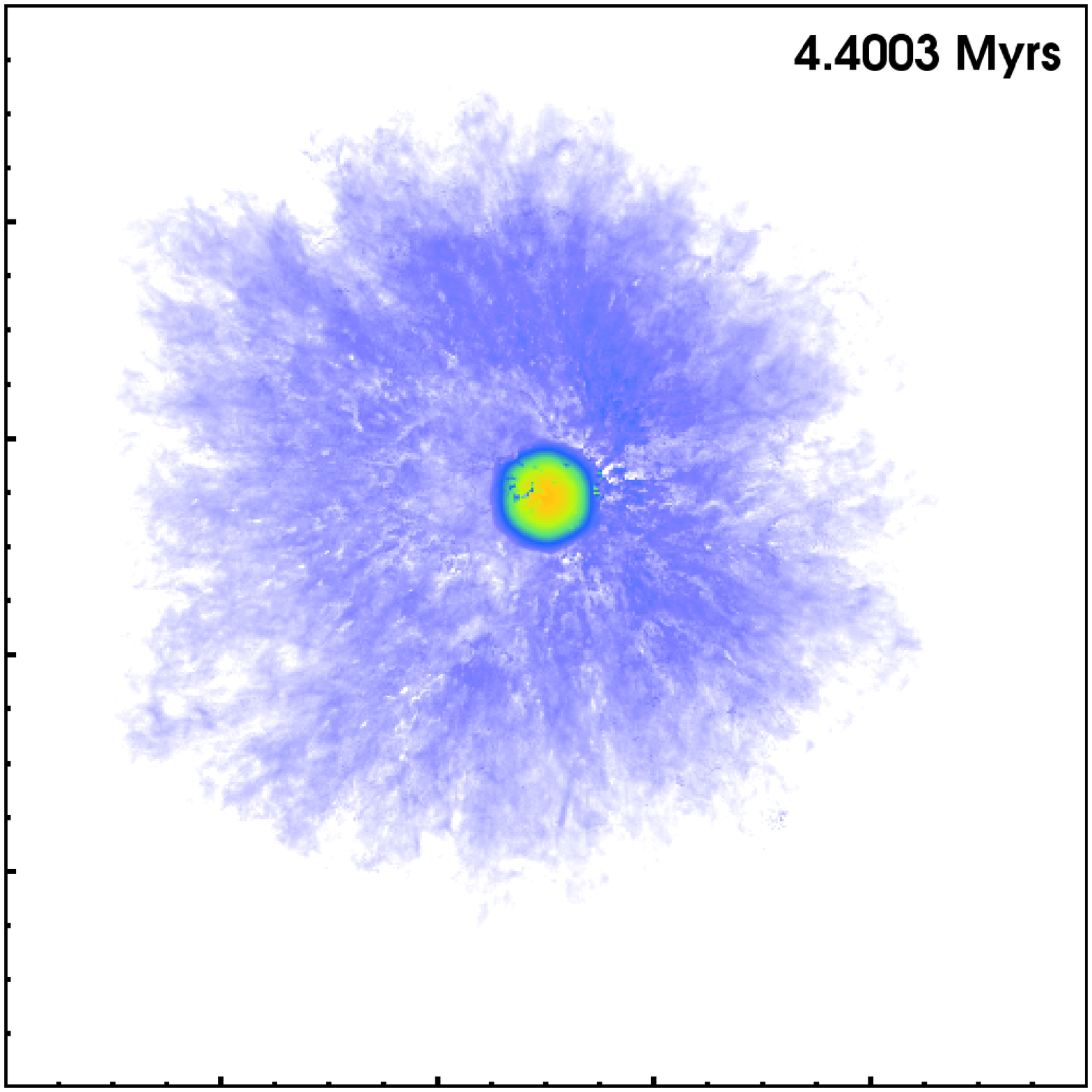}
\includegraphics[width=0.3\textwidth, height=0.3\textwidth]{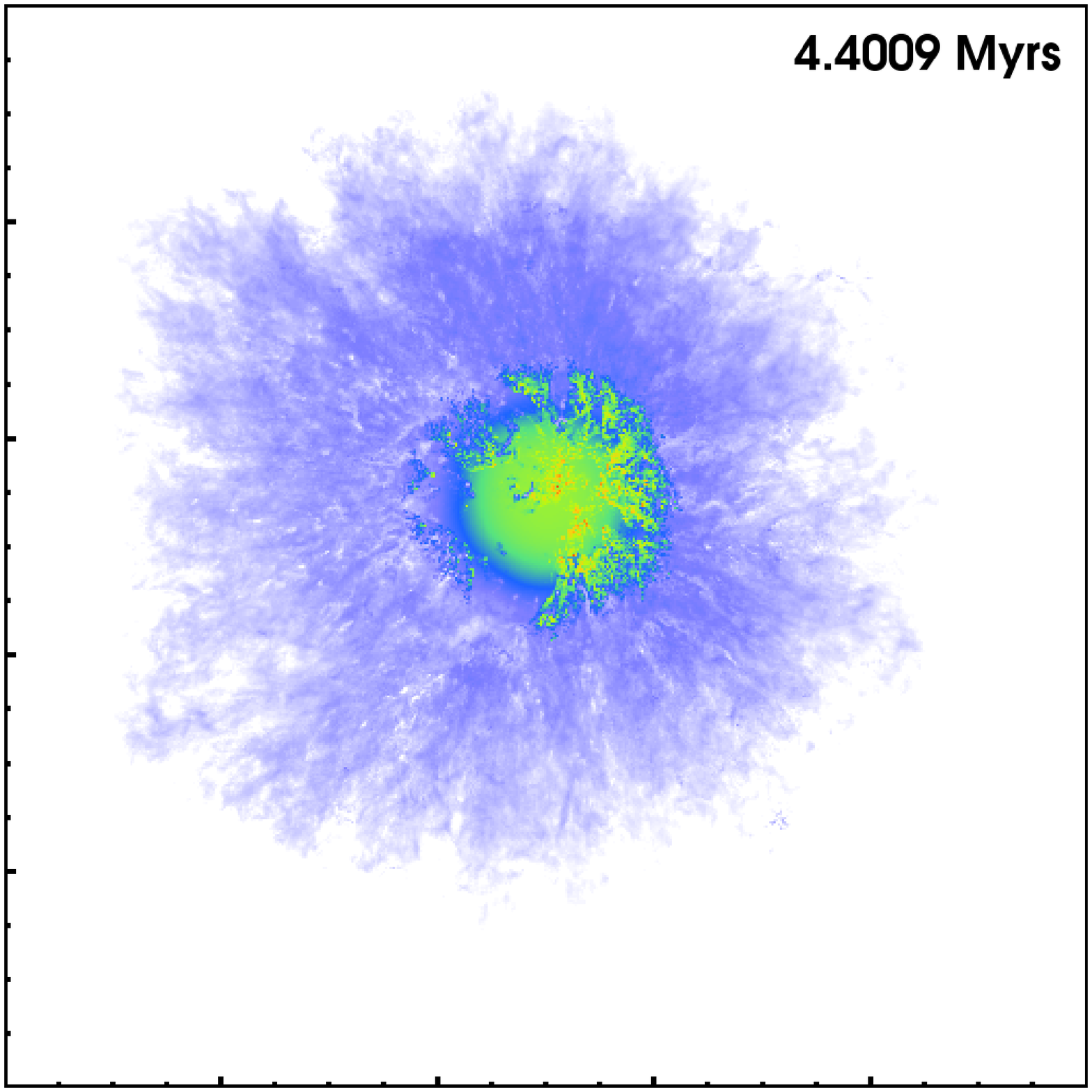}
\includegraphics[width=0.3\textwidth, height=0.3\textwidth]{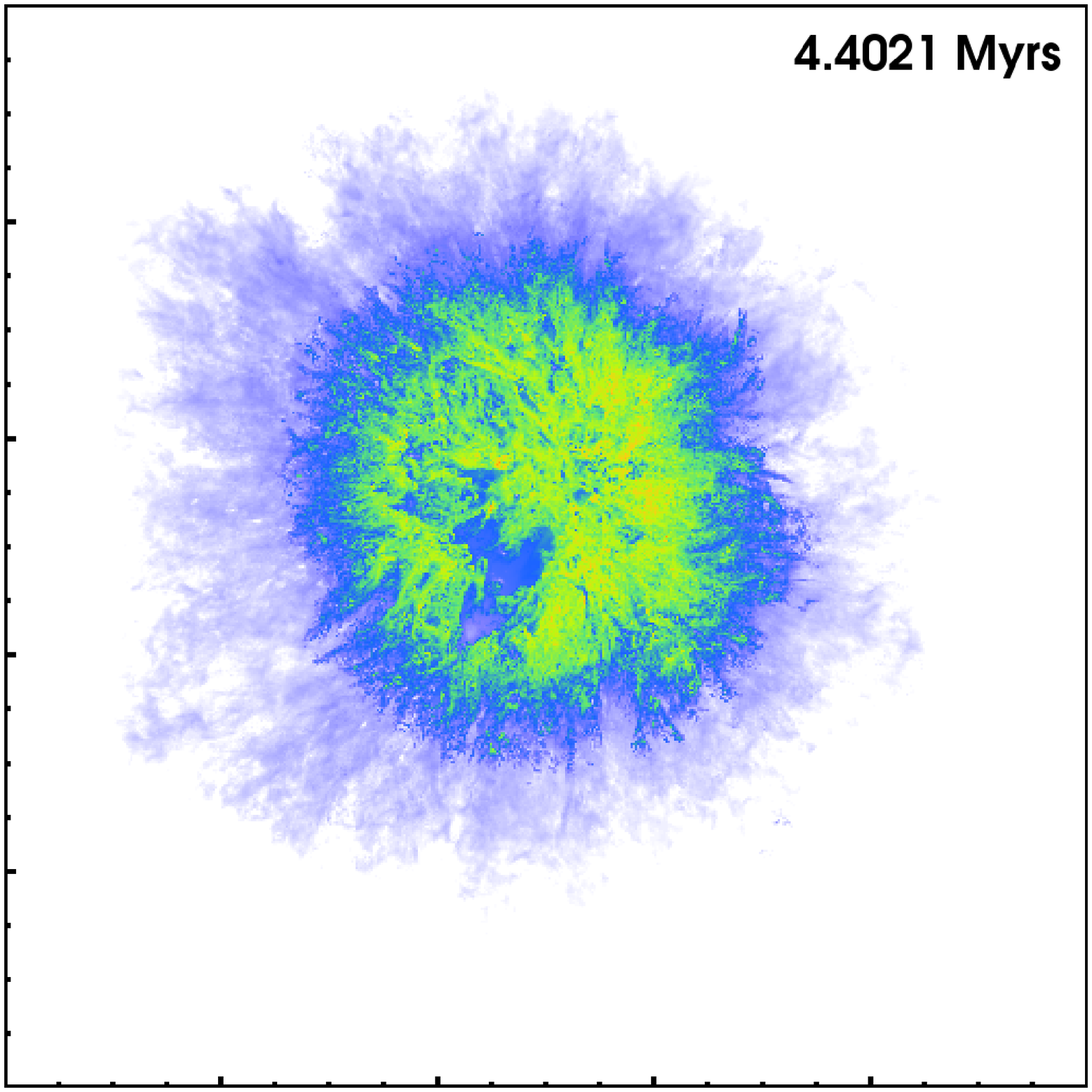}
\includegraphics[width=0.3\textwidth, height=0.3\textwidth]{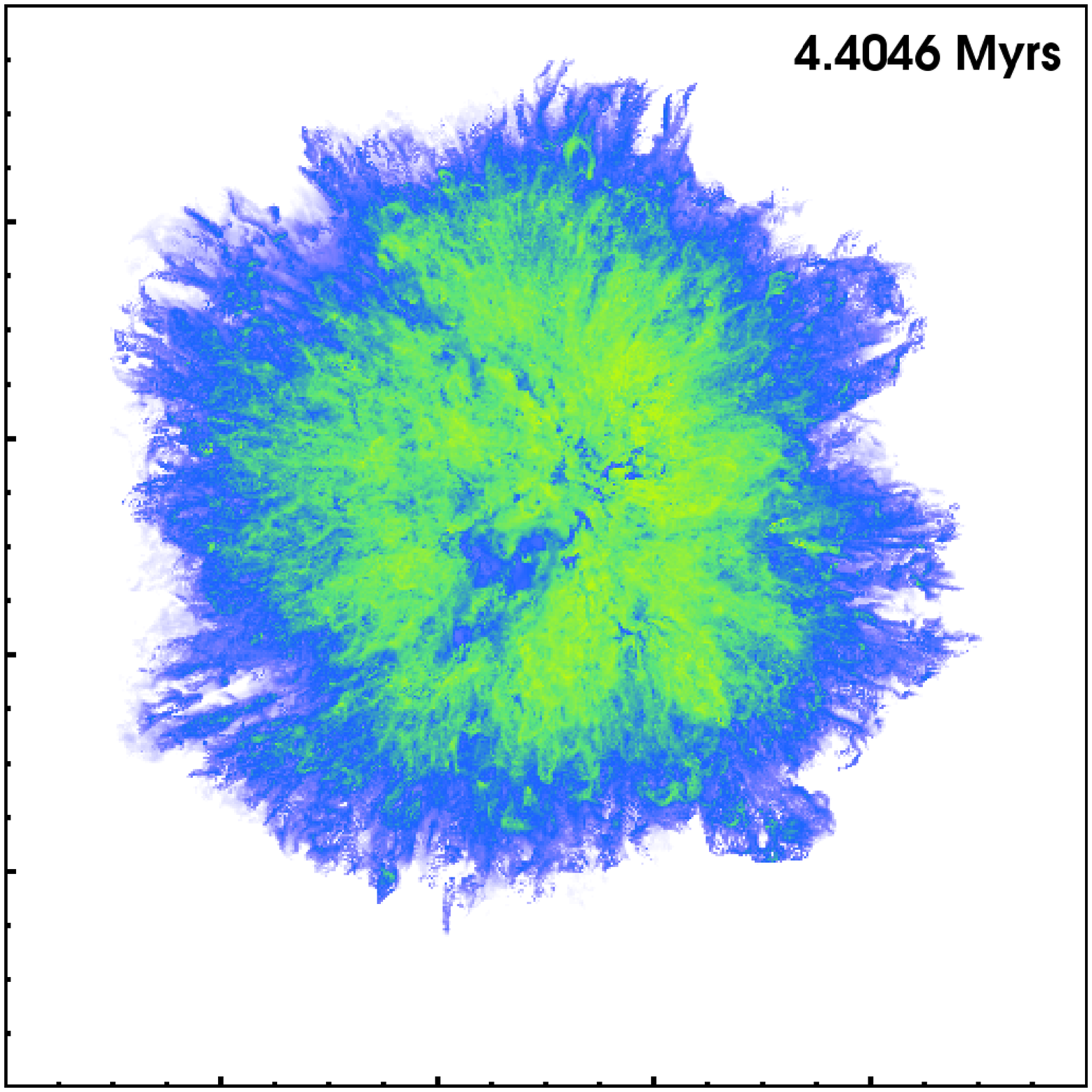}
\caption[Synthetic X-ray image in the BB2 (0.5--2.5\,keV) region during the first 4600\,yrs after the most massive star explodes.]{Same as Fig.~\ref{bb1_sne1}, but for the BB2 (0.5--2.5\,keV) energy band. \label{bb2_sne1}}
\end{figure*}

\begin{figure*}
\centering
\includegraphics[width=0.3\textwidth, height=0.3\textwidth]{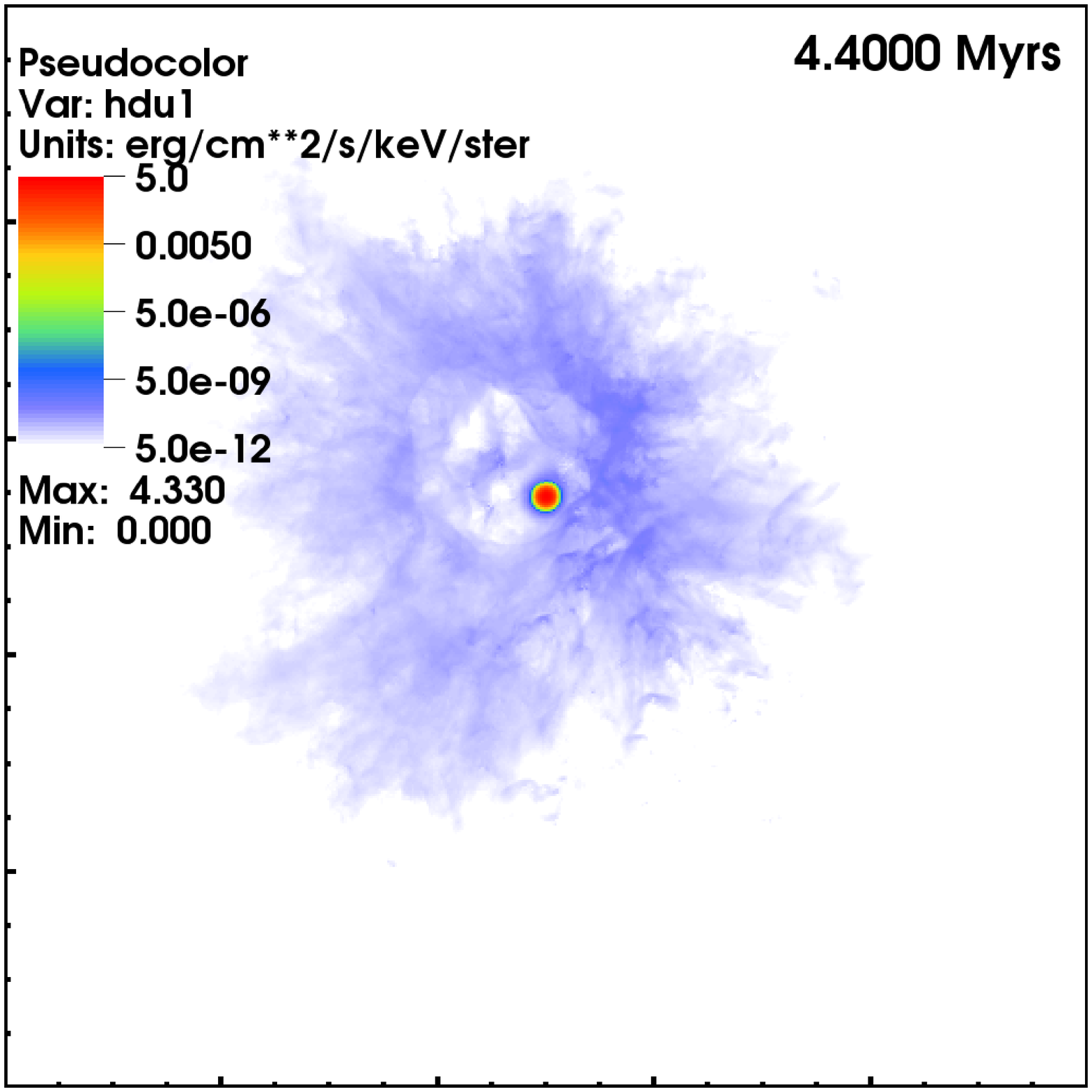}
\includegraphics[width=0.3\textwidth, height=0.3\textwidth]{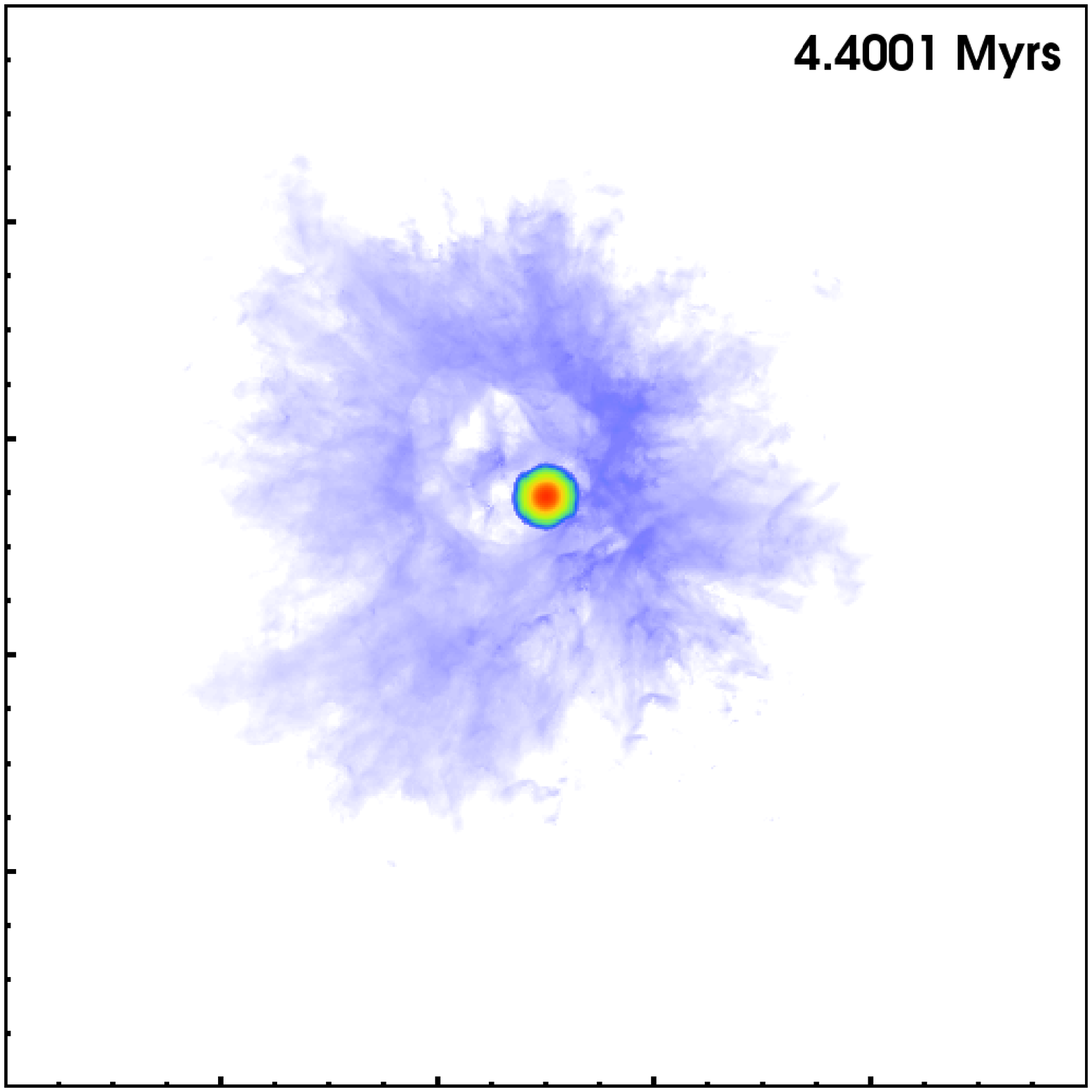}
\includegraphics[width=0.3\textwidth, height=0.3\textwidth]{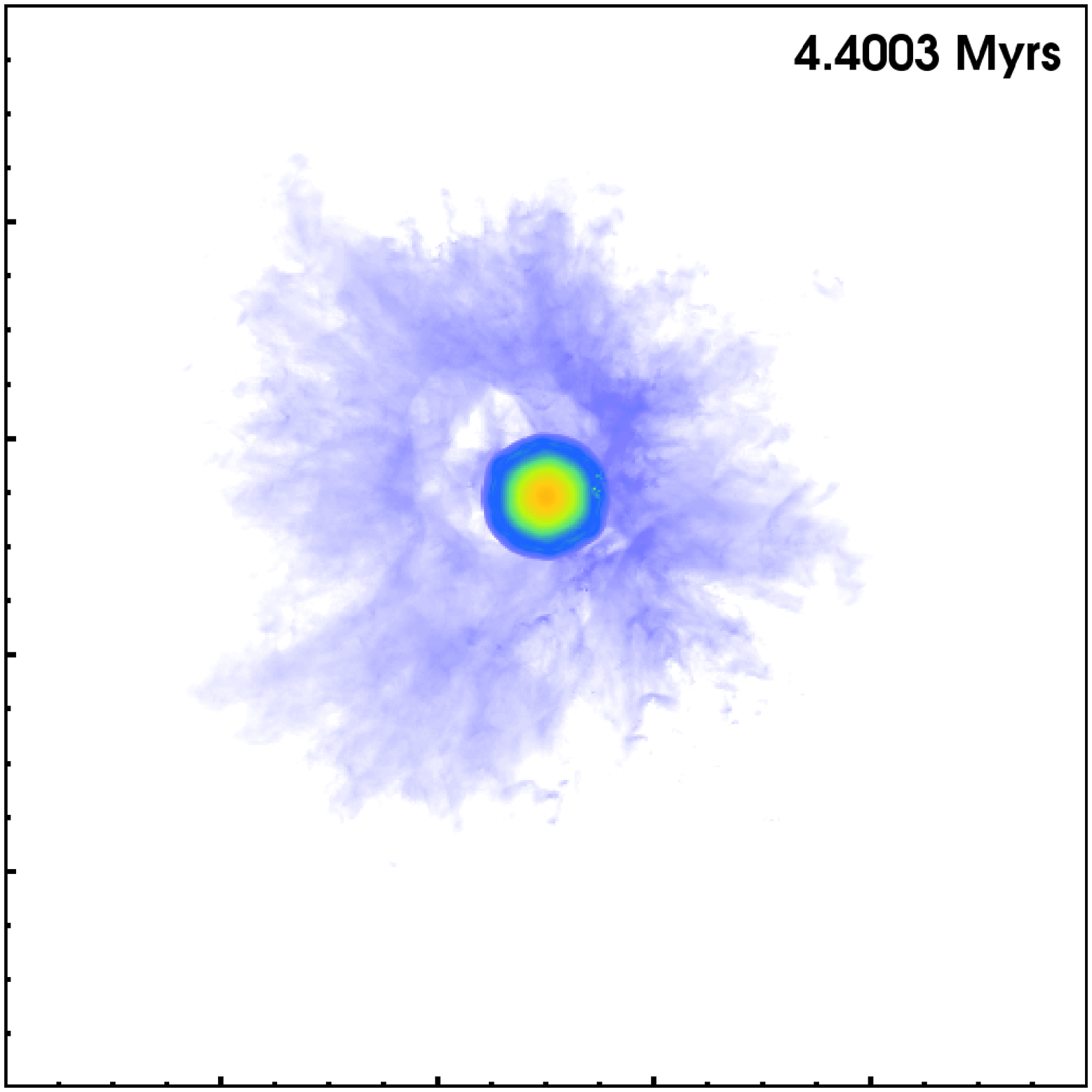}
\includegraphics[width=0.3\textwidth, height=0.3\textwidth]{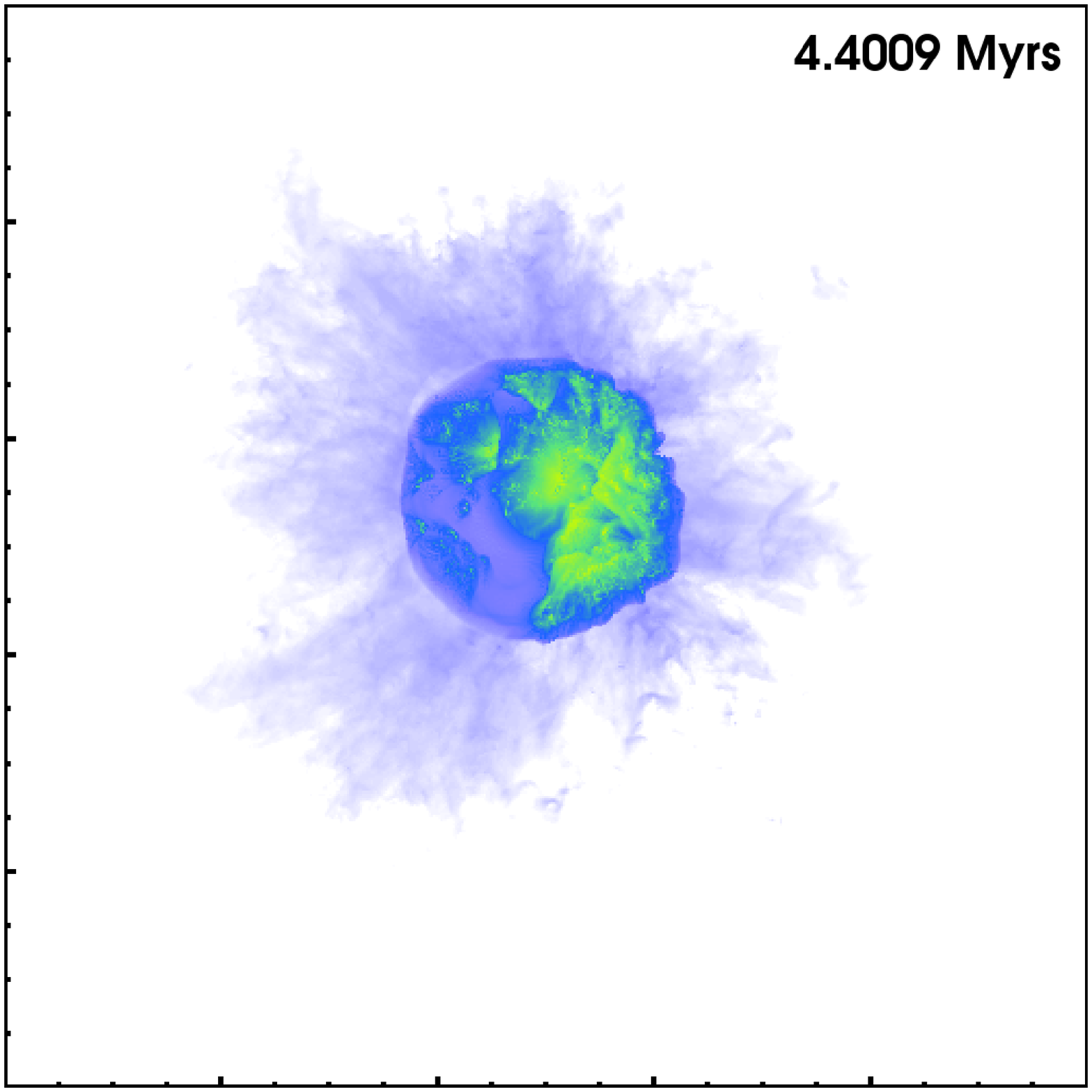}
\includegraphics[width=0.3\textwidth, height=0.3\textwidth]{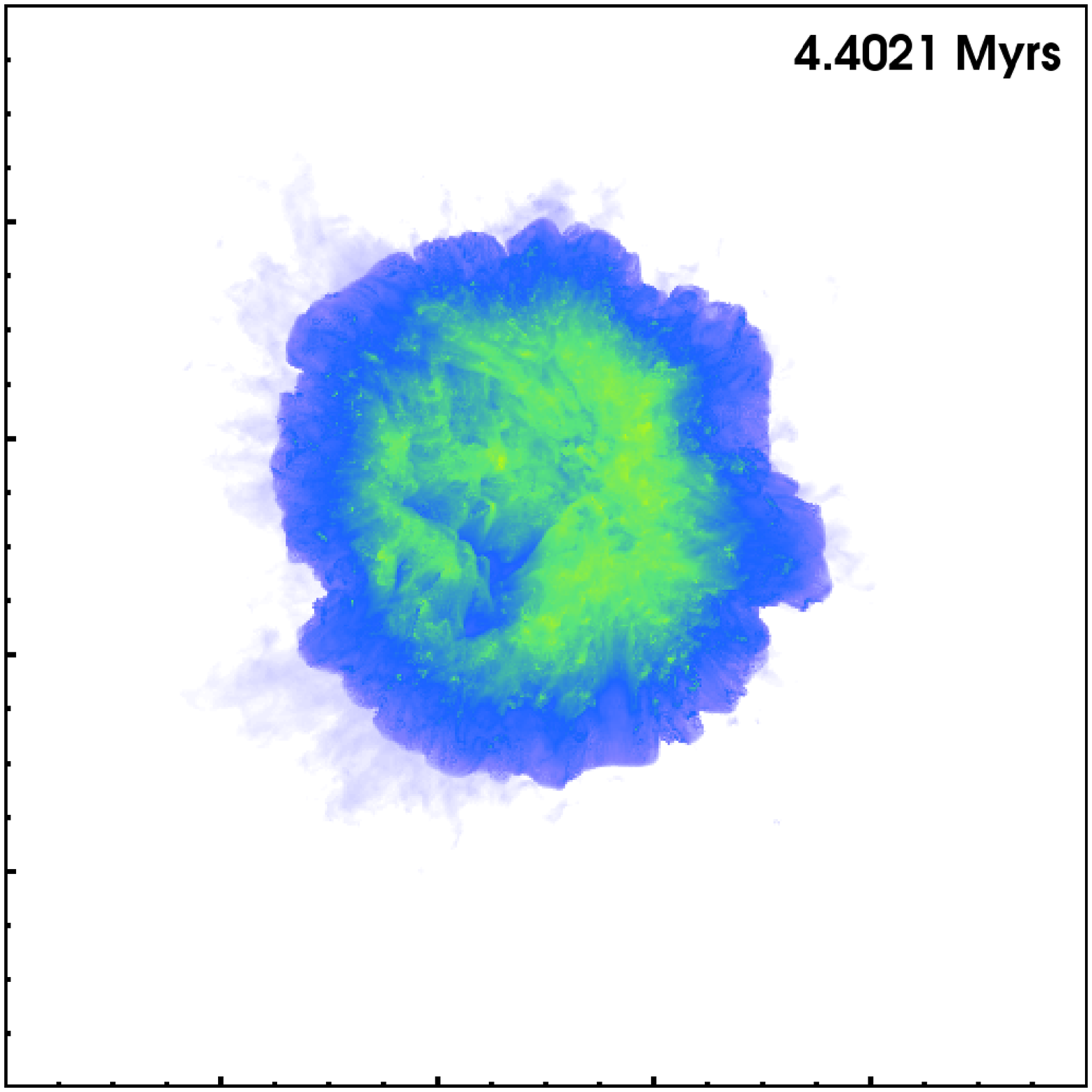}
\includegraphics[width=0.3\textwidth, height=0.3\textwidth]{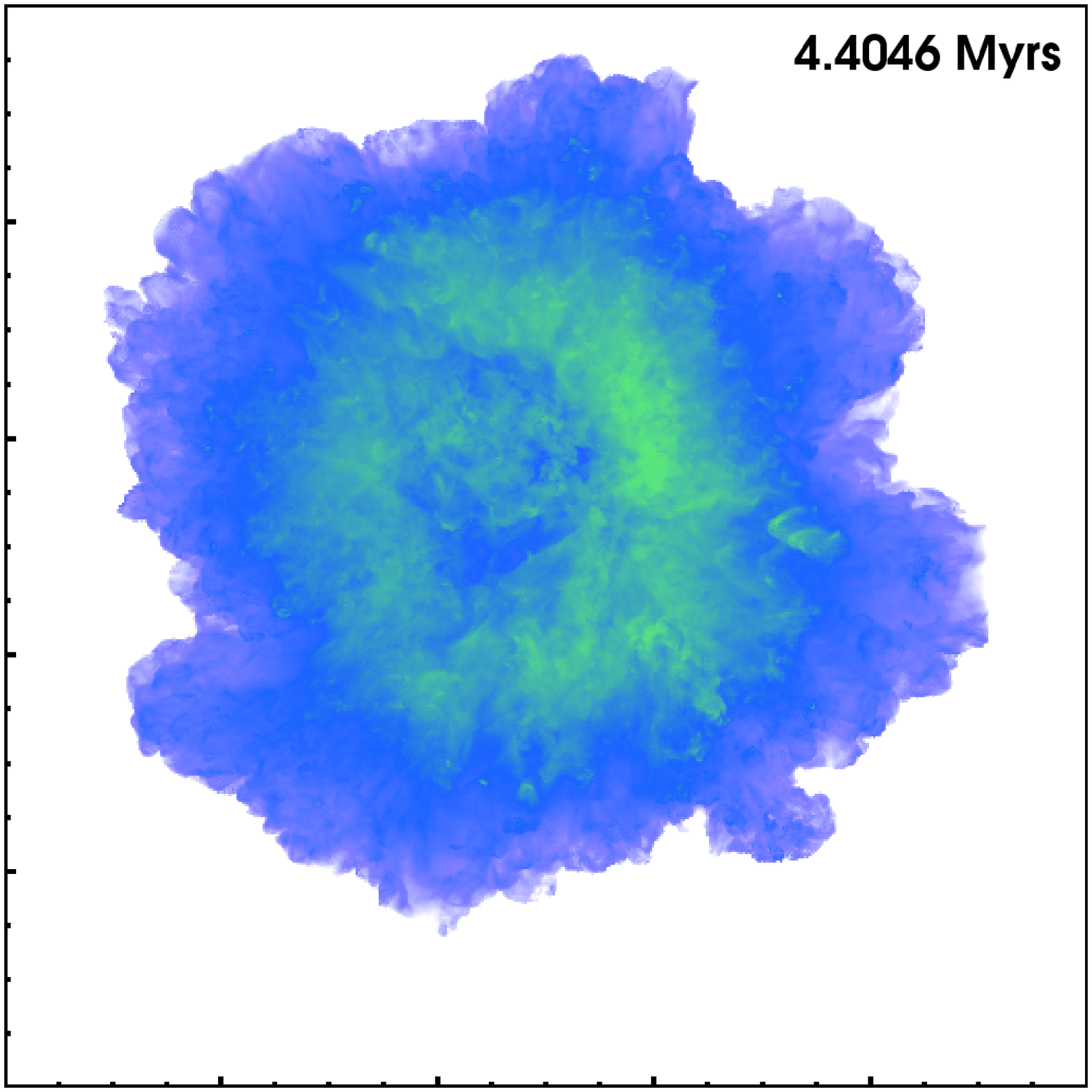}
\caption[Synthetic X-ray image in the BB3 (2.5--10.0\,keV) region during the first 4600\,yrs after the most massive star explodes.]{Same as Fig.~\ref{bb1_sne1}, but for the BB3 (2.5--10.0\,keV) energy band. \label{bb3_sne1}}
\end{figure*}

Synthetic X-ray images in all three energy bands during the explosion are shown in Figs.~\ref{bb1_sne1}-\ref{bb3_sne1}.  When the star first explodes there is high intensity emission at the centre of the cluster, as seen in all three figures, which is a consequence of the implementation of the explosion with thermal rather than kinetic energy.  As the hot ejecta expands outwards the intensity at the centre decreases, although it is still much higher than the pre-explosion levels.

Once the shockwave has expanded out far enough it begins to interact with the high density remains of the GMC clump.  This is apparent from around t\,=\,4.4005\,Myrs onwards.  Prior to this, absorption by dense clumps projected in front of the blast wave is visible (see for example the top middle and top right panels in Fig.~\ref{bb1_sne1}).  As the shockwave sweeps through the inhomogeneous environment successive bowshocks form around each dense cloud that it encounters.  Individual bowshocks can be identified during the first 900\,yrs after the explosion, but at later times these merge to create a single, though highly structured, region of emission with variable surface brightness.  The simulated X-ray emission should be largely unaffected by the explosion setup once the emission from the bow shocks becomes dominant, which as is apparent from the previous discussion of the light curve occurs approximately 900\,yrs after the explosion.

The X-ray image is broadly spherical overall, though there is substantial curvature to the main shock front on local scales.  Ejecta begins to leave the grid approximately 4600\,yrs after the explosion.  By 2000\,yrs after the explosion the most intense emission is observed someway behind the main shock front and the remnant takes on a ``shell-like'' morphology.  This is likely due to the fact that at the time of the SN explosion the densest clouds surrrounding the cluster tend to occur in a shell with inner and outer radii of $\approx$\,5-10\,pc.  Virtually all of the dense gas previously within 5\,pc has been cleared out by the cluster wind, being either ablated and entrained into the cluster wind or pushed away from the cluster by the ram pressure of the cluster wind up to a typical distance of 10\,pc (see Fig.\,7 in Paper I).  The brightest X-ray emission seems to occur where the SN ejectra interacts with the remaining dense clouds in this shell.  In contrast, the position of the forward shock indicates where the SN ejecta has traversed relatively unimpeded through the surrounding medium.

The X-ray image appears to be smoother at the higher energies of BB3, and much more filamentary in the lower energies of BB1 and BB2, even near the edge of the forward shock front.  The emission in BB1 and BB2 is likely picking up gas at relatively low and intermediate temperatures associated with material from dense clouds which is entrained into and partially mixed with the SN ejecta.  Thus the resulting emission traces to some extent the interfaces associated with this process.  However, this gas will be too cool to radiate strongly at the higher energies of the BB3 image, and instead the BB3 image shows the location of hot, but relatively unmixed ejecta.

The forward shock front also shows structures which could be interpreted as ``blow-outs'' (see for example the protrusion to the lower right of the SNR in the bottom right panel of Fig.~\ref{bb3_sne1}).  While these protrusions may indicate that the remnant is expanding into a region with a lower pre-shock density in this direction, it may also be affected by the structure of the surrounding medium that the shock front has encountered at distances considerably prior to this.

The X-ray emission from the environment external (and prior) to the SNR is also of interest.  The emission at t\,=\,0.13\,Myrs and at t\,=\,2.53\,Myrs has already been displayed in Figs.~\ref{xray_04} and~\ref{xray_80}, and discussed earlier.  However, Figs.~\ref{bb1_sne1}--~\ref{bb3_sne1} further illuminate the filamentary emission and absorption which occurs in the BB1 and BB2 energy bands, and the smoother emission which occurs in BB3.  The volume within the reverse shock surrounding the stellar cluster is devoid of any hot gas and is visible as a deficit of emission in the central regions of Figs.~\ref{bb1_sne1} and~\ref{bb2_sne1}.  The highly structured nature of the reverse shock is directly visible in the top row of panels in Fig.~\ref{bb3_sne1}.  It is the X-ray bright parts of the top left panel in Fig.~\ref{bb3_sne1} which first ``light-up'' as the ejecta expands outwards.  Bowshocks around the closest dense clouds to the cluster are responsible in both instances.


\begin{figure}
\centering
\includegraphics[width=0.49\textwidth,height=0.25\textwidth]{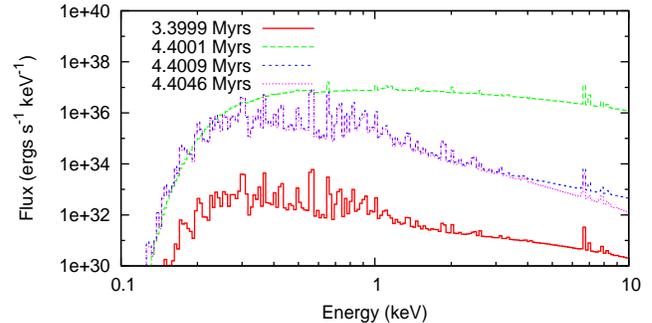}
\caption[The time evolution of the attenuated X-ray spectra of the cluster as the most massive star undergoes a SN explosion.]{The time evolution of the attenuated X-ray spectra of the cluster as the most massive star undergoes a SN explosion.  The solid red line is at a time just before the star explodes, the purple dotted line is $\sim$\,900\,yrs after the explosion and the light blue dot-dashed line is $\sim$\,4,600\,yrs after the explosion. \label{sne_specatt}}
\end{figure}

The time evolution of the attenuated spectra during the period of the first SN explosion is shown in Fig.~\ref{sne_specatt}.  The solid red line is at a time just before the star explodes whilst the light blue dot-dashed line is t\,=\,4600\,yrs after the explosion, which is the approximate time at which the shockwave begins to leave the grid.  The attenuated spectrum for the cluster at t\,=\,4.4009\,Myrs, when bowshock emission begins to dominate, is shown as the purple dotted line in Fig.~\ref{sne_specatt}.  The spectrum is roughly the same shape, albeit considerably more luminous, as that of the pre-SN cluster.  However, whilst the intensity of the soft and medium X-ray energies change little, by the time the ejecta reaches the edge of the grid (light blue dot-dashed line) there is a considerable decrease in the hard X-ray (E$\,\gtrsim$\,3\,keV) emission of the cluster.

\subsection{Further Evolutionary Stages}
After the 35\,M$_{\odot}$ star has exploded the remaining two stars continue in their MS phases for a further 0.1\,Myrs, at which point the 32\,M$_{\odot}$ star begins to follow the same evolutionary path as its predecessor.  At t\,=\,4.5\,Myrs it evolves to a RSG and at t\,=\,4.6\,Myrs it becomes a WR star.  The X-ray lightcurve shown in Fig.~\ref{lightcurve} shows a similar pattern to the previous evolution, in that the X-ray luminosity decreases once the less powerful RSG wind contributes to the cluster wind, while it increases once the star becomes a WR.  However, as the evolution of the 32\,M$_{\odot}$ star occurs so shortly after the first supernova, the luminosity is still declining from the aftermath of that event, partially as a consequence of the blast wave leaving the grid, and so it is hard to distinguish this from the natural decline during the RSG phase.  In fact, the luminosity at this stage is comparable to that of the previous RSG phase despite the loss of one wind source and the commensurate reduction in the momentum and energy flux of the cluster wind.  The 32\,M$_{\odot}$ star explodes at t\,=\,4.9\,Myrs, inparting a further 10\,M$_{\odot}$ of ejecta and 10$^{51}$\,ergs of energy into the simulation.  0.1\,Myrs after this explosion the final remaining star begins the evolutionary sequence already performed by its brethren.

\subsection{Properties of the X-ray Emitting Gas}
The mass, volume and density of the X-ray emitting gas are shown in Table~\ref{xraygasprops}.  The mass of gas with a temperature in excess of 10$^5$\,K increases until t$\,\approx\,0.3$\,Myr when it stands at nearly 7\,M$_{\odot}$.  The mass then drops slightly and stays around 3-4\,M$_{\odot}$ during the remaining MS phase of the cluster wind.  During this time the X-ray emitting volume is just over 50$\%$ of the total simulation volume, since the shocked cluster wind has spread throughout most of the grid.  The average temperature of the X-ray emitting gas is 2.5$\times10^6$\,K, whilst the mass-weighted average is 2.3$\times10^5$\,K, implying that most of the X-ray emitting gas is closer to the 10$^5$\,K mark.  After the first star evolves to the RSG branch the mass of material which is hot enough to produce X-rays decreases along with both the average and the mass-weighted average temperature of the gas.  Once the star evolves further to the WR phase the mass of X-ray producing gas increases by a factor of 10 and the volume of the material at T$\textgreater$\,10$^5$\,K almost doubles.  The average temperature of the X-ray emitting gas increases to 2.8$\times10^6$\,K and the mass-weighted average to 3.2$\times10^5$\,K.

\begin{table*}
\centering
\caption{The mass, density and volume of the X-ray emitting gas, and the average and mass-weighted average temperature of that gas at various times throughout the simulation.}
\begin{tabular}{c c c c c c c c}
\hline
Time & Phase of & Mass at & Density at & Volume at & $\%$ of & log[T$_{\rm av}$] at & Mass-weighted        \\
       & each star & T$\,>10^5$ K  & T$\,>10^5$ K  & T$\,>10^5$ K & Volume at  & T$\,>10^5$        & log[T$_{\rm av}$] at \\
(Myrs) & (35,32,28\,M$_{\odot}$)  & (M$_{\odot}$)  & ($\times10^{-26}\rm\,g\,cm^{-3}$) & ($\times 10^4$\,pc$^3$) & T$\,>10^5$ K    & (K)&T$\,>10^5$ K (K) \\
\hline
\hline
0.00   & MS,MS,MS  & 0.00   &  0.00  & 0.00 &  0.00\,$\%$ & 0.00 & 0.00  \\
0.13   & MS,MS,MS  & 1.89   &  5.17  & 0.25 &  7.63\,$\%$ & 6.40 & 5.58  \\
0.32   & MS,MS,MS  & 6.75   &  3.22  & 1.43 & 43.53\,$\%$ & 6.40 & 5.44  \\
0.63   & MS,MS,MS  & 4.26   &  1.60  & 1.81 & 55.28\,$\%$ & 6.40 & 5.36  \\
0.95   & MS,MS,MS  & 3.25   &  1.20  & 1.83 & 55.92\,$\%$ & 6.40 & 5.36  \\
1.95   & MS,MS,MS  & 3.43   &  1.33  & 1.75 & 53.32\,$\%$ & 6.40 & 5.37  \\
2.53   & MS,MS,MS  & 3.84   &  1.41  & 1.85 & 56.46\,$\%$ & 6.40 & 5.37  \\
3.61   & MS,MS,MS  & 4.06   &  1.42  & 1.95 & 59.40\,$\%$ & 6.40 & 5.39  \\
4.06   & RSG,MS,MS & 2.13   &  1.08  & 1.34 & 40.89\,$\%$ & 6.30 & 5.28  \\
4.31   & WR,MS,MS  & 24.87  &  6.64  & 2.55 & 77.82\,$\%$ & 6.45 & 5.51  \\
4.38   & WR,MS,MS  & 24.99  &  6.69  & 2.54 & 77.54\,$\%$ & 6.40 & 5.51  \\
4.40   & SN,MS,MS  & 37.24  &  9.87  & 2.57 & 78.43\,$\%$ & 6.80 & 6.36  \\
4.4009 & SN,MS,MS  & 51.29  &  13.64 & 2.56 & 78.09\,$\%$ & 6.95 & 5.77  \\
4.404  & SN,MS,MS  & 217.97 &  52.67 & 2.82 & 85.95\,$\%$ & 6.80 & 5.83  \\
4.41   & MS,MS     & 231.57 &  53.04 & 2.97 & 90.64\,$\%$ & 6.65 & 5.86  \\
4.42   & MS,MS     & 132.53 &  29.94 & 3.01 & 91.92\,$\%$ & 6.45 & 5.77  \\
4.56   & RSG,MS    & 1.69   &  2.51  & 0.46 & 14.04\,$\%$ & 6.20 & 5.29  \\
4.78   & WR,MS     & 28.53  &  9.07  & 2.14 & 65.31\,$\%$ & 6.40 & 5.52  \\  
4.90   & SN,MS     & 37.61  &  10.66 & 2.40 & 73.24\,$\%$ & 7.00 & 5.86  \\   
4.94   & MS        & 81.54  &  21.05 & 2.64 & 80.57\,$\%$ & 6.35 & 5.58  \\
\hline
\label{xraygasprops}
\end{tabular}
\end{table*}


\begin{figure}
\centering
\includegraphics[height=0.25\textwidth,width=0.49\textwidth]{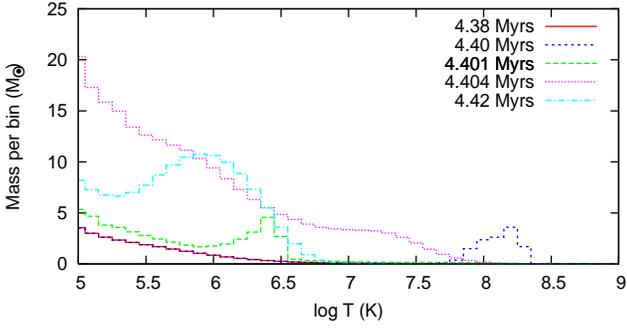}
\caption[Mass of X-ray emitting material above T\,=\,10$^5$\,K at five times during the simulation.]{Mass of X-ray emitting material above T\,=\,10$^5$\,K at five times during the simulation.  The red solid line is at t\,=\,4.38\,Myrs, shortly before the most massive star explodes.  The blue short-dashed line is at t\,=\,4.40\,Myrs immediately after the supernova explosion (this reflects the conditions used to simulate the explosion) and the green long-dashed line is at t\,=\,4.401\,Myrs, 1000\,yrs after the explosion when the emission is dominated by bowshock interactions.  The purple dotted line is at t\,=\,4.404\,Myrs when the ejecta begins to leave the grid.  The light blue dot-dashed line is at t\,=\,4.42\,Myrs, 20,000\,yrs after the explosion.  Each temperature bin is of width 0.1\,dex.  \label{sne_masshist}}
\end{figure}

\begin{figure}
\centering
\includegraphics[height=0.25\textwidth,width=0.49\textwidth]{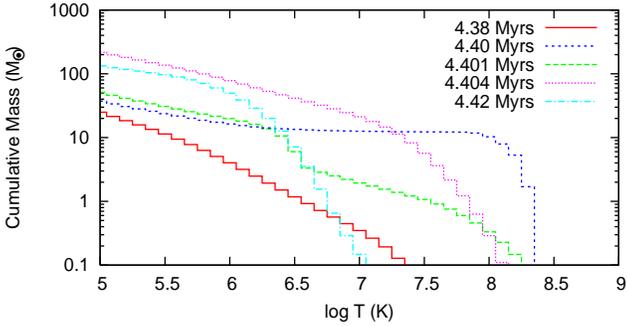}
\caption[The cumulative mass of X-ray emitting material above temperature $T$ at five times throughout the simulation.]{Shows the cumulative mass of X-ray emitting material above T\,=\,10$^5$\,K at five times throughout the simulation.  The red solid line is at t\,=\,4.38\,Myrs, shortly before the most massive star explodes, the blue short-dashed line is at t\,=\,4.40\,Myrs immediately after the SN explosion.  The green long-dashed line is at t\,=\,4.401\,Myrs at which point bowshock emission becomes dominant, and the purple dotted line is at t\,=\,4.404\,Myrs, when SN ejecta begins to leave the grid.  The light blue dot-dashed line is at t\,=\,4.42\,Myrs, 20,000\,yrs after the explosion. \label{decumulative}}
\end{figure}

The mass distribution of the simulation before, during and after the first supernova explosion is shown in Fig.~\ref{sne_masshist}, and the amount of material which is at each temperature above 10$^5$\,K is shown in Fig.~\ref{decumulative}.  The red line shows the temperature distribution of the 25\,M$_{\odot}$ of material above 10$^5$\,K shortly before the explosion at t\,=\,4.38\,Myrs.  There is virtually no gas at temperatures greater than 10$^7$\,K (0.35\,M$_{\odot}$, see the red line in Fig.~\ref{decumulative}), and the mass-weighted average temperature is T$_{av}\,=\,3.2\times 10^5$\,K.  The SN occurs at t\,=\,4.40\,Myrs, and its hot ejecta is visible as the short-dashed blue line in Figs.~\ref{sne_masshist} and~\ref{decumulative}.  As discussed previously, bowshock emission from the SN becomes dominant approximately 900\,yrs after the explosion.  The temperature distribution of the 51\,M$_{\odot}$ of material above 10$^5$\,K at this time is shown by the green long-dashed line in Figs.~\ref{sne_masshist} and~\ref{decumulative}.  There is a small amount of material above 10$^7$\,K ($\sim\,2\rm\,M_{\odot}$), but the majority of the material is between 10$^5-10^{6.5}$\,K, after which there is an obvious decrease in X-ray emitting material.  The mass-weighted average temperature is T$_{av}\,=\,5.9\times 10^5$\,K at this time.

The SN ejecta begins to leave the grid at t\,=\,4.404\,Myrs, shown as the purple dotted line in Figs.~\ref{sne_masshist} and~\ref{decumulative}.  There is approximately 218\,M$_{\odot}$ of material above 10$^5$\,K at this time (see Table~\ref{xraygasprops}), with $\sim\,10\%$ (21\,M$_{\odot}$) of that material above 10$^7$\,K.  20,000\,yrs after the explosion there is again virtually no gas at T\,$\textgreater$\,10$^7$\,K as the shock heated gas slowly cools (shown by the light blue dot-dashed line in Fig.~\ref{sne_masshist}).  However, there is approximately 5 times more material between 10$^5-10^7$\,K than before the supernova, with a peak at about T\,=\,10$^6$\,K.  Fig.~\ref{decumulative} also reveals that the maximum temperature of gas at t\,=\,4.42\,Myrs is actually lower than that at t\,=\,4.38\,Myrs, at T$_{\rm max}\,=\,10^{7.4}$\,K and T$_{\rm max}\,=\,10^{7.7}$\,K respectively.


\section{Comparison to 1D Bubble Models}
\label{sec:comp_theory}
\subsection{Wind-Blown Bubble Models}
\citet{Chu95} derived an analytical expression for the X-ray emission from a \citet{Weaver77} wind-blown bubble (WBB), in terms of various physical parameters which are observable, such as the density and size of the bubble.  The predicted X-ray luminosity in the 0.1--2.4\,keV band is:

\begin{equation}
\centering
\label{eq:chu}
L_{X}= \left(1.1\times10^{35}\rm{\,erg\,s^{-1}}\right) \xi I(\tau)L_{37}^{33/35}n_0^{17/35}t_6^{19/35}
\end{equation}

\noindent where $\xi$ is the metallicity relative to the solar value, L$_{37}$ is the mechanical luminosity of the stellar wind(s) in units of 10$^{37}$\,ergs\,s$^{-1}$, n$_0$ is the number density of the ambient medium in cm$^{-3}$ and t$_6$ is the age of the bubble in 10$^6$\,yr.  The above equation contains a dimensionless temperature $\tau$, where the dimensionless integral I($\tau$)\,=\,(125/33)\,-\,5$\tau^{1/2}$\,+\,(5/3)$\tau^3$\,-\,(5/11)$\tau^{11/2}$ and $\tau\,=\,0.16L_{37}^{-8/35}n_0^{-2/35}t_6^{6/35}$.

At t\,=\,0.3\,Myrs the expected luminosity in the 0.1--2.4\,keV energy band as predicted using Equation~\ref{eq:chu} is $\rm{L_{X}}\approx\,2.06\,\times\,10^{35}\,\rm{ergs\,s^{-1}}$ using the average ambient density of the mostly intact GMC clump of n$_0\,\approx\,\rm{250}\,cm^{-3}$.  This compares to the combined luminosity from our BB1 and BB2 energy bands, which at $\rm{L_{X}}=\,3.9\,\times\,10^{31}\,\rm{ergs\,s^{-1}}$ is roughly 5000 times lower than the prediction from the standard \citet{Weaver77} bubble.  Because the edge of the bubble expands off the grid at t$\,\sim\,0.2\,$Myrs, we will have somewhat underestimated the true luminosity of our simulation, but it is clear that a large discrepancy nevertheless remains.

By t\,=\,2.53\,Myrs the destruction of the GMC clump is well advanced and we should clearly reduce our estimate of the appropriate value for the ambient density, n$_0$.  Using an estimate of the density as n$_0\,\approx\,\rm{0.3}\,cm^{-3}$ (which is just 50$\%$ greater than the low density medium which surrounds the GMC clump in the simulations), the predicted X-ray luminosity from Equation~\ref{eq:chu} would be $\rm{L_{X}}\approx\,1.4\,\times\,10^{34}\,\rm{ergs\,s^{-1}}$.   Although our simulation only ``captures'' a small proportion of the X-ray luminosity at this time since a lot of the hot gas has flowed through the grid boundaries, the estimate from Equation~\ref{eq:chu} is approximately 2000 times larger than the luminosity from our simulations at this time.  Since this factor is likely to be many times greater than the ``true'' X-ray luminosity from our simulation (i.e. the luminosity we would infer if our grid were big enough to contain the expanding bubble), we conclude that Equation~\ref{eq:chu} consistently overestimates the X-ray luminosity produced from our simulations by a large margin. 

\citet{Harper-Clark09} also provide an analytical expression for the expected X-ray luminosity of a Weaver wind-blown-bubble:

\begin{align}
\label{eq:hc+m}
\nonumber L_{X} \sim 3\times10^{38}\xi\left(\frac{L_{w}}{4\times10^{38}\rm{erg\,s^{-1}}}\right)^2 &\left(\frac{20pc}{r}\right)^3\left(\frac{6\times10^6\,K}{T}\right)^2 \\
 &\cdot\left(\frac{3.6\times10^6\,\rm{yr}}{t}\right)^2
\end{align}

\noindent where they have assumed an X-ray cooling rate $\Lambda_{X}\,\approx\,3\times10^{-23}\xi\,\rm{erg\,s^{-1}\,cm^{3}}$ and $t$ is the age of the cluster/wind source.  Applying this expression to the Carina nebula overestimates the observed luminosity by a factor of 10$^4$ \citep{Harper-Clark09}.  In the following we apply Equation~\ref{eq:hc+m} to our simulated cluster at a time when the stars are on their MS phases and the mechanical luminosity of the cluster wind is L$_w\,=\,1.16\times10^{36}\rm\,ergs\,s^{-1}$.  The average temperature of the X-ray emitting gas at t\,=\,2.53\,Myrs is 2.5$\times10^6$\,K (see Table~\ref{xraygasprops}).  We only capture a small part of the bubble volume at this time.  However, we can be guided by what an observer may choose as the bubble radius.  If the ISM column to the cluster was substantially higher than our assumed value of 10$^{21}\rm\,cm^{-2}$, the emission below 2.5\,keV may be almost completely absorbed, in which case Fig.~\ref{xray_80} shows that only the harder emission might be detected.  The radius that an observer might then infer for the ``bubble'' in the BB3 image in Fig.~\ref{xray_80} could then be approximated to 6\,pc.  This leads to a predicted X-ray luminosity of L$_X\,\approx\,10^{36}\rm\,ergs\,s^{-1}$ using Equation~\ref{eq:hc+m}, and an overestimate of the intrinsic emission ``captured'' in our simulation by roughly 4 orders of magnitude.  To bring the values from Equation~\ref{eq:hc+m} in line with the simulated results would require a bubble radius of 130\,pc.


Clearly some of the underlying assumptions made in these equations are incompatible with the simulated results.  The two main assumptions in the \citeauthor{Weaver77} bubble model which differ from our simulations are that the energy deposited by the stellar winds is confined within the bubble and that the surrounding ISM is homogeneous.  As discussed in both \citet{Harper-Clark09} and Paper I, leakage of hot gas from the bubble interior leads to a significant reduction in the pressure.  This in turn reduces the X-ray luminosity so that it is well below that from a confined bubble \citep[c.f.][]{Harper-Clark09}.  It is interesting that the calculated X-ray luminosity from our simulations is roughly 3--4\,dex lower than the predictions from the confined bubble model, which is of order of the same difference between observations of real clusters and the confined bubble model.


\subsection{Cluster Wind Models}
\citet{Chevalier85} and \citet{Canto00} derived an analytical model describing the cluster wind flow that results from the multiple interactions of the stellar winds produced by the stars of a dense cluster of massive stars.  \citet{Rodriguez07} developed the work of \citet{Canto00} to include a non-uniform stellar distribution.

\begin{figure}
\centering
\includegraphics[height=0.25\textwidth, width=0.49\textwidth]{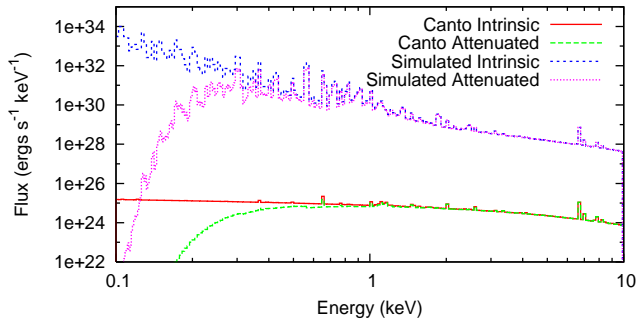}
\caption{A comparison between the intrinsic and attenuated X-ray spectrum of our simulated cluster whilst all three stars are on the MS with the \citet{Canto00} analytical cluster wind model.  The red solid line shows the intrinsic and the green long-dashed line shows the attenuated X-ray luminosity as calculated from the \citeauthor{Canto00} model.}
\label{ms_spectra}
\end{figure}

In order to compare these models with the simulations presented in this work, the cluster mass-loss rate and the average velocity were set to $\rm{\dot M}_{\rm cl} = 9\times10^{-5}\rm\,M_{\odot}\,yr^{-1}$ and $\rm v_{\rm cl} = 2000\rm\,km\,s^{-1}$ respectively, equivalent to the simulated cluster at a time when all three of the stars are still on the MS (see Table~\ref{evolution} for the individual stellar properties).  The cluster radius in both of the analytical models was set to $\rm R_{\rm cl} = 0.04\rm\,pc$, which is consistent with stellar clusters of comparable mass.  The X-ray spectrum of these solutions are shown in Fig.~\ref{ms_spectra}, along with the X-ray spectrum of our simulated cluster at t\,=\,2.53\,Myrs.


It is clear that the \citet{Canto00} model produces X-ray luminosities significantly lower than from our simulated cluster.  This is expected considering the lesser degree of confinement of the stellar winds inherent in the cluster wind model \citep[c.f.][]{Harper-Clark09}.



\section{Comparisons to Observations}
\label{sec:comp_obs}

\subsection{Young Massive Stellar Clusters}

It is widely observed that there is a deficit of X-ray emission from stellar clusters compared with predictions based on the WBB model of \citet{Weaver77} \citep{Dorland86, Dorland87, Oey96, Rauw02, Dunne03, Smith05, Harper-Clark09}.  Many explanations of this effect have been suggested, for example lower stellar luminosities, mass or energy loss from the bubble, or highly efficient mass loading which reduces the temperature of the cluster below X-ray temperatures.  However, mass-loading may also produce higher X-ray luminosities \citep{Stevens03}.  Section~\ref{sec:comp_theory} demonstrated that the X-ray luminosities produced by the model in Paper I also exhibit a lower luminosity than predicted by WBB models.  A comparison will now be made between our model results and observations of M17 and the Rosette Nebula.  For a full literature review of young massive stellar clusters from which diffuse X-ray emission has been detected see Table~\ref{stellarclustercomp} and Appendix\,A.

\begin{table*}
\centering
\caption[The properties of young massive stellar clusters from which diffuse X-ray emission has been detected.]{The properties of young massive stellar clusters from which diffuse X-ray emission has been detected.  The clusters are ordered roughly by age.  The abbreviation ``CF'' in column 4 is short for champagne flow.  Further details and references for values in this table can be found in Appendix\,A.}
\begin{tabular}{c c c c c c c c }
\hline
Cluster/Region  & Age           & Distance & X-ray       & Thermal/NT & L$_X$             & kT            & N$_H$       \\
Name            & (Myr)         & (kpc)    & Morphology  &            & (erg\,s$^{-1}$)   & (keV)         & (cm$^{-2}$) \\
\hline
\hline
RCW 38          & $\lesssim$ 1 & 1.7           & CF/blowouts & NT         & 3$\times10^{32}$   & 0.2            & 9.5$\times10^{21}$ \\
Omega (M 17)    & $\sim$\,0.5   & 2.0           & CF/blowout & T          & 2$\times10^{34}$   & 0.28,0.29,0.57 & 1.6,5,10$\times10^{21}$ \\
Westerlund 2    & $\lesssim$ 2 & 2.85$\pm$0.43 & CF          & T	   & 4.6$\times10^{33}$ & 0.1,0.8,3.1    & 4,12,12$\times10^{21}$ \\
Rosette         & 2             & 1.55          & CF          & T          & 7$\times10^{32}$   & 0.06,0.8       & 2$\times10^{21}$ \\
Hourglass       & 1--2.5        & 1.3           & CF          & T          & $\lesssim6.6\times10^{32}$ & 0.63  & 1.1$\times10^{22}$ \\
Arches          & 2--2.5        & 8             & CF          & T (+NT)    & 3.8$\times10^{33}$  & 2.56          & 1.1$\times10^{23}$ \\
NGC 2024 (Flame)& 0.3--3        & 0.415         & CF          & T          & 2$\times10^{31}$    & 11            & 0.21,3.3$\times10^{22}$ \\
Orion (M 42)    & 3             & 0.49          & CF   & T          & 5.5$\times10^{31}$  & 0.17          & 4.1($\textless$ 1.0)$\times10^{22}$ \\
Quintuplet      & 3.5--4        & 8             & CF          & T          & 3$\times10^{33}$    & 10$^{+4.6}_{-2.7}$ & 3.8$\times10^{22}$ \\
NGC 3603        & 1--4          & 7$\pm$1       & CF          & T          & 2.6$\times10^{35}$  & 0.53          & 2$\times10^{22}$ \\
Westerlund 1    & 4--5          & 4--5          & CF          & T (+NT?)   & 1.7--30$\times10^{33}$ & 0.7,3.0    & 2$\times10^{22}$ \\
NGC 3576N       &               & 2.8$\pm$0.3   & CF          & T (+NT)    & 5.9$\times10^{33}$   & 0.11,0.5,0.67 & 0.3,1.3$\times10^{22}$ \\
NGC 3576S       &               & 2.8$\pm$0.3   & Blowout     & T          & 1.1$\times10^{34}$   & 0.31,0.53    & 1.3,0.3$\times10^{22}$ \\
\hline
\end{tabular}
\label{stellarclustercomp}
\end{table*}

\subsubsection{M17}
M17 is a young blister HII region located on the northeast edge of one of the largest GMCs in the Galaxy, at an approximate distance of 1.55\,kpc.  It is estimated to be only $\sim\,$0.5\,Myrs old \citep{Chini08,Hoffmeister08}.  M17 is photoionized by the massive stellar cluster NGC\,6618, within which \citet{Broos07} have identified 14 O-stars.  An earlier study by \citet{Hanson97} identified at least 9 O-stars and a few late-O/early-B stars.

The diffuse X-ray emission from M17 has previously been analyzed by \citet{Dunne03}, \citet{Townsley03}, \citet{Hyodo08}, and most recently by \citet{Townsley11}.  The total X-ray luminosity is thought to be L$_X\,=\,2\times10^{34}\rm\,ergs\,s^{-1}$.  At t\,=\,0.44\,Myrs, the simulated cluster has an intrinsic 0.1--10\,keV luminosity (BB1+BB2+BB3) of L$_X\,=\,7.48\times10^{32}\rm\,ergs\,s^{-1}$, approximately 25 times lower than in M17.  However, given the number and type of O-stars in M17, the difference in emission from our simulated cluster and the observed X-ray luminosity from M17 can be considered to be perfectly acceptable.  \citet{Townsley03} estimate the mass of plasma at $T\,\sim\,10^6$\,K to be 0.3\,M$_{\odot}$.  Whilst the simulated cluster has 4.26\,M$_{\odot}$ at t\,=\,0.63\,Myrs (see Table~\ref{xraygasprops}), this includes gas at temperatures greater than $T\,\textgreater\,10^5$\,K.  The amount of material at temperatures above $10^6$\,K in our simulated cluster is actually 0.31\,M$_{\odot}$ at this time, which is remarkably similar to the observations of M17.  For more information on M17 see Appendix\,A.

\subsubsection{Rosette Nebula}
The Rosette Nebula is a blister HII region at the tip of the giant Rosette molecular cloud.  It is estimated to be $\sim\,$2\,Myrs old \citep{Hensberge00}.  A study carried out by \citet{Martins12} identified 7 O-stars in the cluster NGC\,2244 contained within the Rosette Nebula.  The earliest spectral type so far detected is O4V((f)).

\citet{Townsley03} find that soft diffuse X-ray plasma surrounds the OB association and fills the nebula cavity completely.  This plasma likely originates from the O star winds and is later brought to thermalization by wind-wind interactions or by shocking against surrounding molecular material.  They estimate the mass of the diffuse plasma at T$\,\textgreater 10^6$\,K to be $M_X\,\sim\,0.05\rm\,M_{\odot}$, which is again much lower than in our simulated cluster at this time (3.43\,M$_{\odot}$, see Table~\ref{xraygasprops}).  However, the amount of X-ray emitting material above T$\,\textgreater\,10^6$\,K in our simulated cluster is actually 0.3\,M$_{\odot}$, and above T$\,\textgreater\,10^7$\,K  it is 0.02\,M$_{\odot}$.  These values are a much better match to the observations of the Rosette Nebula.

 The intrinsic 0.5--2\,keV luminosity is $\approx\,7\times10^{32}\rm\,ergs\,s^{-1}$, with no significant emission detected above 2\,keV.  At t\,=\,1.96\,Myrs, our simulated cluster has an intrinsic 0.5--2.5\,keV luminosity (BB2) of L$_X\,=\,6.46\times10^{30}\rm\,ergs\,s^{-1}$.  Given the higher mass-loss rate of the Rosette cluster \citep[$\dot M_*\,=\,2.5\times10^{-6}\rm\,M_{\odot}\,yr^{-1}$,][]{Stevens03} and the higher number of O-stars present, this is a reasonably close match to the simulations.  For more information on the Rosette Nebula see Appendix\,A.





\subsection{Young SNRs from Core-Collapse SNe}
When the stars in the simulation explode they input 10\,M$_{\odot}$ of material and 10$^{51}$\,ergs of energy into the surroundings.  As seen in Fig.~\ref{snelightcurve} the results of this explosion should only be trusted after 900\,yrs when the dominant source of X-ray emission is from interactions between the blast wave and the surrounding clumpy medium.  The ejecta begins to leave the grid after 4600\,years, at which point hot gas and its corresponding emission begins to be lost.  A comparison will now be made between the model and observations of young SNRs which are of age $900 < \rm t < 4600$ years.  Four SNRs matching these criteria are identified, as noted in Table~\ref{snecomparison}.  Each of these are now discussed in turn.

\begin{table*}
\centering
\caption{Young SNRs from core-collapse SNe compared with the simulated results.  The SNRs are ordered roughly by age.  References can be cound in the accompanying text in \S5.2.  Five sets of simulation results are given, spanning the estimated age range of the observed SNRs.}
\begin{tabular}{c c c c c c c c c}
\hline
SNR  & Alternative     & L$_x$   & Diameter & Age  & V$_{\rm exp}$        & Temperature & Distance & Prog.      \\
Name & Name            &(ergs$\rm\,s^{-1}$)  & (pc) & (yr) &  (km$\rm\,s^{-1}$) &(keV)        & (kpc)     & Mass (M$_{\odot}$) \\
\hline
\hline
1E0102.2-7219        &    & $\sim\,8.8\times10^{35}$& $\sim$\,12    & $\sim$\,1000--2100 &  1000            & 2.5--4.5,0.4--1  &  60       & 32     \\
MSH\,11-54 & G292.0+1.8    & $\textgreater\,7.2\times10^{32}$ & 15   & 2700--3200 & $\lesssim$\,1200         & 1.05,0.37   & $\textless$\,6 & 20-40  \\
N\,132D     & SNR\,0525-69.6 & 4.5-7.5$\times10^{37}$ & $\sim$\,20--25 & 3150       &  2250--3700              & 0.6--0.7    & 55             & 30-35  \\
Puppis\,A             &    & 1.2$\times10^{37}$     & $\sim$\,32     & 3700--4450 & $\textgreater$\,1500     & 0.6         & 1.3-2.2  & $\textgreater\,$25 \\
\hline
Simulation           & & 2.71$\times10^{37}$    & $\sim$\,18   & 1060       & $\sim$\,5900     & 0.68        & 1           & 35                 \\
Simulation           & & 2.49$\times10^{37}$    & $\sim$\,27   & 2060       & $\sim$\,5200     & 0.68        & 1           & 35                 \\
Simulation           & & 1.67$\times10^{37}$    & $\sim$\,29   & 2630       & $\sim$\,1600     & 0.68        & 1           & 35                 \\
Simulation           & & 1.05$\times10^{37}$    & $\sim$\,30   & 3300       & $\sim$\,1000     & 0.54        & 1           & 35                 \\
Simulation           & & 1.02$\times10^{37}$    & $\sim$\,31   & 3860       & $\sim$\,1000     & 0.54        & 1           & 35                 \\
\hline
\label{snecomparison}
\end{tabular}
\end{table*}

\subsubsection{1E0102.2-7219}
1E0102.2-7219 (hereafter 1E0102) is a SNR in the Small Magellanic Cloud (SMC) with an inferred kinematic age of $\sim$\,2100\,yrs \citep{Eriksen01} at a distance of $\sim$\,60\,kpc \citep{Wada13}.  The progenitor is thought to have been a Wolf-Rayet star with a zero-age MS (ZAMS) mass of $\sim$32\,M$_{\odot}$ that underwent significant mass loss prior to exploding as a Type Ib/c or IIL/b supernova.  \citet{Gaetz00} used \emph{Chandra} to image the SNR, finding it to be almost ``textbook'', with a hotter outer ring surrounding a cooler, denser inner ring which is likely the reverse-shocked stellar ejecta.  The diameter of the SNR was estimated to be 40'' by \citet{Hughes88} (approximately 12\,pc at 60\,kpc distance).  More recently, \citet{Hughes00b} estimated the radius of the blast wave to be 6.4\,pc, in good agreement with their earlier work.  \citet{Hughes00b} also estimated an expansion age of $\sim$\,1000\,yrs, though \citet{Eriksen01} disagree, deriving a free expansion age of 2100\,yrs.   However, recently \citet{Wada13} have proposed that the source is a Be/NS binary.  There is little interstellar extinction along the line of sight to 1E0102 which allows a comprehensive, multi-wavelength analysis from the X-ray to the radio domain.

\citet{Hughes00b} estimated the expansion velocity of the blast wave to be 6000\,km\,s$^{-1}$ based on a radius of 6\,pc and an age of 1000\,yrs, although \citet{Flanagan04} find from Doppler shifts that the majority of the bulk matter is moving at a lower 1000\,km\,s$^{-1}$.  \citet{Hughes00b} also estimate the temperature in the postshock region to be 0.4--1.0\,keV.  \citet{Gaetz00} estimated the upper limit on X-ray emission of the central source to be $\textless\,9\times10^{33}\rm\,ergs\,s^{-1}$, whilst \citet{Wada13} estimate the 0.5--10\,keV luminosity of the Be/NS binary to be $\sim\,8.8\times10^{35}\rm\,ergs\,s^{-1}$ using Suzaku data.  This is only a factor of 2 lower than our intrinsic 0.5--10\,keV X-ray emission from the simulated cluster, which is L$_X\,=\,1.53\times10^{36}\rm\,ergs\,s^{-1}$ and L$_X\,=\,1.69\times10^{36}\rm\,ergs\,s^{-1}$ for 1000 and 2000\,yrs after the first SN explosion respectively.  As the simulation assumes collisional ionization equilibrium (CIE), which is unlikely to be the case in 1E0102, it is not surprising that higher luminosities are obtained from the simulation.  The clumpy environment in the simulated cluster may also be partially responsible, if the environment of 1E0102 has lower density and/or is more homogeneous.  The expansion velocity of the simulated blastwave is between $\sim$\,6000--8000\,km\,s$^{-1}$ (see Table~\ref{snecomparison}), which although high is similar to the estimate of the blast wave by \citet{Hughes00b}.

\subsubsection{MSH 11-54}
Also known as G292.0+1.8, this is a core-collapse SN with an estimated age of 2700 - 3200\,years \citep{Chevalier05,Winkler09,Tanaka13}, and a distance of 6\,kpc \citep{Gaensler03}.  It is one of only a handfull of O-rich SNRs known today \citep{Park07,Ghavamian12}.  The X-ray emission from such O-rich SNRs is thought to arise from faster, non-radiative shocks in lower density ejecta and interstellar gas.  The central source is thought to be a pulsar wind nebula.  The SNR has a radius of approximately 15\,pc assuming a distance of 6.2\,kpc \citep{Gaensler03}.

\citet{Gonzalez03} derived an average temperature for the SNR using two components - a high temperature plasma associated with the supernova blast wave and a low temperature plasma from the reverse shock.  These two components were estimated to be 1.05 $\pm$ 0.34 and 0.37 $\pm$ 0.18\,keV respectively.  The progenitor star is estimated to have had a mass of 30-40\,M$_{\odot}$, though \citet{Hughes94} estimated a lower mass of 20-25\,M$_{\odot}$.  A more recent estimate of the temperature by \citet{Park07} using \emph{Chandra} data found a highly non-uniform distribution of hot, X-ray emitting gas in the remnant ranging from kT$\sim$5\,keV in the NW regions to around kT$\sim$0.7\,keV in the SE. These results are a promising match with our results 2000--3300\,years after the explosion of the 35\,M$_{\odot}$ star, when the average temperature of the X-ray emitting plasma is $\simeq\,$0.54--0.68\,keV.  The SN explosion producing MSH\,11-54 is thought to have been asymmetric, which would explain the spatial variation of temperatures and the greater expansion of the remnant towards the NW.  \citet{Hughes03} found the unabsorbed 0.3--10\,keV X-ray luminosity of the central pulsar to be L$_X\,=\,7.2\times10^{32}\rm\,ergs\,s^{-1}$, but made no estimate for the entire remnant.  Our simulated cluster at around this time has a total 0.1--10\,keV intrinsic luminosity of L$_X\,\sim\,1\times10^{37}\rm\,ergs\,s^{-1}$.

\subsubsection{N132D}
N132D is one of the brightest SNRs in the Large Magellanic Cloud (LMC) and has an estimated age of 3150\,years \citep{Morse95} and an inferred progenitor mass of 30-35\,M$_{\odot}$ \citep{Blair00}.  With a diameter of 80'', the distance to the SNR of approximately 55\,kpc \citep{Hughes87} implies a real diameter of $\sim$\,21\,pc.  This is a similar estimate to the extent of the X-ray shell, which has an estimated radius of 12\,pc \citep{Morse95}.  The expansion velocity of the SNR has been estimated by several authors \citep[e.g.][]{Morse95,Hwang92}, with values ranging from 2250--3700\,km\,s$^{-1}$.  In comparison, the radius of our simulated SNR at around 3300\,yrs after the explosion is $\sim\,15$\,pc, with an inferred expansion velocity of $\sim$\,4400\,km\,s$^{-1}$. 

 The X-ray luminosity in the 0.2--4\,keV energy band was estimated by \citet{Hughes87} to be 4.5--7.5$\times10^{37}\rm\,ergs\,s^{-1}$, based on thermal plasma temperatures of 10$^{6.8}-10^{7.1}$\,K and a hydrogen column density of 10$^{21}-10^{21.5}\rm\,cm^{-2}$ \citep{Raymond77}.  The estimated luminosity is actually higher than the simulated results (L$_X\,=\,1.05\times10^{37}\rm\,ergs\,s^{-1}$ at 3300\,yrs), despite the fact that the simulations assume CIE.  However, given the inherrent differences in the simulated and actual environments these results can be considered a reasonable match.  The plasma temperature is very similar to that found in the other SNRs mentioned, at approximately 0.6-0.7\,keV, compared with an average of 0.54\,keV from our simulated remnant (see Table~\ref{xraygasprops}).  There is patchy X-ray absorption around the remnant thought to be caused by gas just outside the molecular cloud towards the nothern tip of N132D \citep{Kim03}.

\subsubsection{Puppis A}
Puppis A is a nearby Galactic SNR and has age estimates ranging from 3700\,yr \citep{Winkler85} to 4450\,yr \citep{Becker12}, making it most comparable to the simualation 3900\,yrs after the explosion (see Table~\ref{snecomparison}).  A distance of 2.2\,kpc has been estimated based on HI and CO studies \citep{Reynoso03}, although a closer distance of 1.3\,kpc has also been proposed by \citet{Woermann00} based on OH line detections.  This remnant is embedded in a complex region composed of large atomic and molecular clouds and an interstellar density gradient.  The remnant is about 50' in diameter (approximately 32\,pc at a distance of 2.2\,kpc).  A progenitor mass of 25$\rm\,M_{\odot}$ was inferred by \citet{Canizares81}.

Optical knots detected from Puppis A are evident only in the northeast, implying the ejection of the matter during the explosion was asymmetric \citep{Katsuda08}.  Oxygen-rich filaments are detected to have radial velocities higher than $\sim$\,1500\,km\,s$^{-1}$.  These filaments are interpreted as SN ejecta which have remained mostly uncontaminated by the ISM \citep{Winkler85}.


Recently, \citet{Dubner13} studied Puppis A using \emph{Chandra} and XMM-Newton.  They estimated the X-ray luminosity between 0.3 and 8.0\,keV to be L$_X\,=\,1.2\times10^{37}\rm\,ergs\,s^{-1}$ assuming a distance of 2.2\,kpc.  The X-ray emission from Puppis A appears to be dominated by the swept-up ISM due to very low metal abundances \citep{Hwang08}.  The total intrinsic X-ray luminosity of our simulated remant 3860\,yrs after the explosion is L$_X\,=\,1.02\times10^{37}\rm\,ergs\,s^{-1}$, which is a reasonable match to Puppis A.

The average temperature in the remnant is 0.6\,keV, very similar to the average temperature of 0.54\,keV seen in the simulated remnant at 3860\,years after the explosion (see Table~\ref{xraygasprops}).

\section{Conclusion}
\label{sec:xray_conclusions}
This paper investigates the X-ray emission from a massive young stellar cluster embedded in an inhomogeneous GMC clump treating only the mechanical effects from winds and supernovae.  The hydrodynamical input model was previously simulated in Paper I, and this work explores the emission arising from that model.  Initially the dense parts of the clump decrease the observed X-ray emission due to attenuation, but once the cluster wind has destroyed and ablated a large portion of this material the attenuation from the ISM material is dominant.

At very early times, when the wind material is still confined by the inhomogeneous GMC material the X-ray luminosity is reasonably bright, at L$_X \approx 5\times 10^{33}\rm\,ergs\,s^{-1}$.  However, as the cluster wind errodes and destroys the surrounding clump it is no longer completely confined and therefore hot gas is able to leak through the gaps in the shell.  This causes a reduction in the X-ray luminosity as the pressure within the bubble decreases.  Once the low density gas from the clump has been ablated away the covering factor of the cluster remains more or less constant, leading to an approximately constant intrinsic X-ray luminosity of 1.7$\times 10^{32}\rm\,ergs\,s^{-1}$ and an attenuated X-ray luminosity of 7$\times 10^{30}\rm\,ergs\,s^{-1}$.

The most massive star becomes a RSG at t\,=\,4.0\,Myrs, resulting in a large drop in the X-ray luminosity in all three of the X-ray broadbands studied.  The most dramatic decrease is seen in the BB3 (2.5--10.0\,keV) emission, where the attenuated X-ray luminosity drops four orders of magnitude, from L$_X \sim 2\times 10^{29}\rm\,ergs\,s^{-1}$ to L$_X \sim 2\times 10^{25}\rm\,ergs\,s^{-1}$ by the end of the RSG phase.  The drop in X-ray luminosity in the other two broadbands over this period is around a factor of 50 and 100 for BB1 (0.1--0.5\,keV) and BB2 (0/5--2.5\,keV) respectively.  Although a lot of material is deposited in the RSG-enhanced cluster wind, the amount of material at X-ray emitting temperatures is very low, contributing to the lack of X-ray emission observed at this time. 

100,000\,years later the most massive star further evolves to become a WR star, causing a dramatic increase in X-ray emission in all three broadband regions studied.  The amount of material at a temperature greater than 10$^5\rm\,K$ increases by an order of magnitude over that seen in the RSG stage, and a total of 78$\%$ of the computational volume contains X-ray emitting material.  The high momentum cluster wind sweeps up the slower moving material deposited in the previous phase and heats it to high temperatures, with the average temperature of hot gas ($T\,\textgreater\,10^5\,$K) at this time being around T$=2.5\times 10^6$\,K.  The total attenuated X-ray emission increases to L$_X \sim 5\times 10^{33}\rm\,ergs\,s^{-1}$, which is about 30 times greater than that observed when all three stars were on the MS.

At t\,=\,4.4\,Myrs the most massive star in our simulation explodes as a SN, ejecting 10$\rm\,M_{\odot}$ of material and 10$^{51}\rm\,ergs$ of energy into the simulation.  Due to the way in which the explosion is initialised, the emission from the SNR only becomes independent of the initial conditions of the explosion once the interaction of the blastwave with the surrounding material becomes dominant.  In this work this occurs approximately 900\,years after the explosion.  The ejecta begins to leave the grid 4600\,years after the explosion, and therefore the emission from the SNR can be compared with observations of SNRs only between the ages of $900 < \rm t < 4600$ years.

The supernova of the 35\,M$_{\odot}$ star was compared with four young core-collapse SNe with ages ranging from $\sim$\,1000--4450\,yrs.  Although collisional ionization equilibrium was assumed in our simulation, which is unlikely to be true for such young remnants, we find that the X-ray luminosity and electron temperatures are reasonable matches to observational results reported in the literature.  Unfortunately, as the ejecta begins to leave the grid 4,600\,yrs after the initial explosion no comparisons can be made with older remnants.

The simulated emission from the cluster during the wind-dominated phases is substantially lower than predicted by 1D spherically symmetric WWB calculations, but is higher than predicted by cluster wind models.  The fact that our simulations fall between these two models ties in very nicely with the theory of leaky bubble models, where the WBB is only partially confined.  The simulated results match reasonably well to actual observations of several massive young stellar clusters.  This is likely to be due the assumptions made in these calculations being overly simplified compared to the simulated model.  Firstly, the assumption that the hot wind material is confined within a bubble is very much not the case, and a reduction in the pressure in the simulated cluster leads to a reduction in the X-ray luminosity.  The surrounding density is not homogeneous as described in these models, which will lead to local areas of confinement and leakage.  Clearly the highly complex environment of young massive star forming regions requires similar complexity in simulations in order to better understand their properties.

\section*{Acknowledgements}

HR acknowledges a Henry Ellison Scholarship from The University of Leeds and JMP acknowledges funding from the Royal Society for a University Research Fellowship and from the STFC.  We would also like to thank the referee for their timely and useful comments.

\appendix
\section{Notes on Individual Stellar Clusters}
\label{sec:app1}

\subsection{RCW\,38}
RCW\,38 is a very young ($< 1$\,Myr), highly embedded ($A_{\rm V} \sim 10$), and close ($D=1.7\,$kpc) stellar cluster 
surrounding a central pair of O5.5 stars (IRS\,2) \citep[e.g.][]{Winston11}. The dominant IRS\,2 stars have cleared a region completely free of dust out to a radius of 0.1\,pc. The extinction is patchy, and the HII region appears to be breaking out of the surrounding molecular gas in some directions. Extended warm dust is found throughout a $2-3$\,pc region and coincides with extended ($1.25 \times 1.75\,$pc) X-ray plasma which is predominantly non-thermal \citep{Wolk02}. The power-law index of the emission steepens toward the cluster core. Contamination of the diffuse emission by unresolved point-sources is not significant at distances of more than 0.15\,pc ($\sim 15$\,arcseconds) from the cluster center, though may be responsible for the more thermal nature of the diffuse emission measured in the core \citep{Wolk06}. The cause of the non-thermal emission remains unclear.

The diffuse emission is strongest in the central core near IRS\,2, and is confined on the southeast along a ridge. Recent \emph{Spitzer} observations reveal that winds from IRS\,2 have caused outflows towards the northeast, northwest and southwest of the central cluster \citep[see e.g. Fig.~4 in][]{Winston12}.

An excellent review of this cluster is given in the Handbook of Star Formation, where the luminosity of the diffuse X-ray emission is given as about $3\times10^{32}\,{\rm ergs\,s^{-1}}$ \citep{Wolk08}.

\subsection{The Omega Nebula (M17)}
M17 is a very young blister H II region located on the northeast edge of one of the largest giant molecular clouds in the Galaxy, with an extent of $4^{\circ}$ \citep[$\sim 110$\,pc,][]{Elmegreen79}. The geometry of M17 is thought to resemble the Orion Nebula HII region except that it is seen edge-on rather than face-on \citep{Meixner92,Pellegrini07}. M17 is photoionized by the massive stellar cluster NGC\,6618, which has 14 identified O stars \citep{Broos07}, and is estimated to be $\sim 0.5$\,Myr old \citep{Chini08,Hoffmeister08}. The distance has recently been determined using trigonometric parallax to be 2.0\,kpc \citep{Xu11}.
Several obscured O4-O5 stars form a central 1 arcminute ring in NGC\,6618, and are principally responsible for ionizing the nebula. Extinction is patchy \citep[$A_V = 3-15$ with an average of 8 to the OB stars, though some parts of the cluster have $A_V > 20$,][]{Hanson97}. The earliest O stars are an O4+O4 visual binary known as Kleinmann's Anonymous Star \citep{Kleinmann73}, which may in fact be a pair of colliding wind binaries \citep{Broos07,Hoffmeister08,Rodriguez12}. Evidence for an older ($2-5$\,Myr) stellar population to the North is presented by \citet{Povich09}.

The diffuse X-ray emisison from M17 has previously been analyzed by \citet{Dunne03}, \citet{Townsley03}, \citet{Hyodo08}, and most recently by \citet{Townsley11}.  It has relatively high surface brightness and blows out to the east of the cluster, extending nearly 10\,pc from the cluster. The plasma appears to be channelled by the famous northern and southern ionization bars \citep{Povich07}, and maintains roughly constant temperature as it flows \citep{Townsley03,Hyodo08}. Although obscuration changes across the field, a global fit to the diffuse emission with a 3-temperature NEI model yields $kT_{1}= 0.28$, $kT_{2} = 0.29$ and $kT_{3} = 0.57$\,keV, with the highest temperature component providing 56\% of the intrinsic luminosity \citep{Townsley11}. The absorption to each of these emission components increases with the temperature of the component, so that the $kT_{3}$ component suffers 6 times as much obscuration as $kT_{1}$. There are indications that the shocked gas is not in complete ionization equilibrium, which is suggestive of it recently being shocked. Several gaussian lines are also needed - the cause is speculated to be charge exchange processes. The total X-ray luminosity is $2.0\times10^{34}\,{\rm erg\,s^{-1}}$.

\citet{Townsley03} previously determined that the X-ray plasma had a mass of $0.15\,M_{\odot}$, which when rescaled to a distance of 2.1\,kpc becomes $0.3\,M_{\odot}$ (for an assumed distance $D$, $V_{\rm x} \propto D^{3}$, $L_{\rm x} \propto D^{2}$, $n_{\rm e,x} \propto (L_{\rm x}/V_{\rm x})^{1/2} \propto D^{-1/2}$, and $M_{\rm x} \propto n_{\rm e,x}V_{\rm x} \propto D^{5/2}$). \citet{Townsley03} determine that $n_{\rm e,x} \sim 0.3\,{\rm cm^{-3}}$. The analysis by \citet{Hyodo08}, which does not quite include the most easterly extent of the plasma, is generally consistent with the earlier work of \citet{Townsley03}, except for the determination of a significantly lower plasma temperature of $\approx 0.25$\,keV.

\subsection{Westerlund\,2 (RCW\,49)}
Westerlund\,2 (hereafter W2) is a compact young open cluster embedded in and responsible for the luminous HII region RCW\,49. 
W2 contains at least a dozen OB stars. Two WR stars, WR20a and (especially) WR20b, lie outside the cluster core \citep[see references in][]{Churchwell04}.  W2 is also located in the direction of one of the Galaxy’s strongest sources of $\gamma$-rays \citep{Aharonian07,HESS11}. The distance to W2 has been very difficult to pin down, with estimates ranging from 2 to more than 8\,kpc in the literature, but a new study puts it at $2.85\pm0.43$\,kpc, and determines an age of no more than 2\,Myr \citep{Carraro13}.

Diffuse X-ray emission from W2 was identified in a \emph{Chandra} observation \citep{Townsley05}. The emission is brightest at the core of W2, and extends preferentially towards the west. The emission can be fitted with a 3-temperature thermal plasma model with $kT_{1}= 0.1$, $kT_{2} = 0.8$ and $kT_{3} = 3.1$\,keV, with the highest temperature component providing 30\% of the intrinsic luminosity \citep{Townsley05}. Assuming a distance of 2.3\,kpc, the absorption corrected 0.5-8\,keV luminosity is $L_{\rm x} = 3\times10^{33}\,{\rm ergs\,s^{-1}}$. At $D=2.85$\,kpc, this increases to $L_{\rm x} = 4.6\times10^{33}\,{\rm ergs\,s^{-1}}$. The absorbing column to $kT_{1}$ is less than the identical columns to $kT_{2}$ and $kT_{3}$. The hardest thermal component is not well constrained, and replacing it with a power-law component ($\Gamma=2.3$) also gives an acceptable fit. The diffuse flux will be slightly underestimated due to the use of a 5\,arcmin radius extraction region and a nearby on-chip background region. 

More recently diffuse emission has also been analyzed from a \emph{Suzaku} observation \citep{Fujita09}.  The \emph{Chandra} pointing was used to determine the point source contamination to the \emph{Suzaku}-derived diffuse emission, and the central region ($r \textless 2$\,arcmin) was masked out. The diffuse emission is found to extend to an 8 arcminute radius. The spectral analysis is broadly consistent with the earlier results of \citet{Townsley05}.

\subsection{The Orion Extended Nebula (M42)}
The Orion Nebula Cluster (ONC), also known as the Trapezium cluster, contains the nearest rich and concentrated sample of pre-MS stars. The OB members of the ONC photoionize the Orion Nebula (M\,42), a blister HII region at the near edge of Orion A, the nearest giant molecular cloud ($D \approx 450$\,pc). \citet{Gudel08} recently detected diffuse, soft ($0.3-1$\,keV) X-ray emission in the Extended Orion Nebula (EON). The characteristic temperature of the plasma is $kT \approx 0.2$\,keV.  The intrinsic X-ray luminosity in the $0.1-10$\,keV energy band is $L_{\rm x} = 5.5\times10^{31}\,{\rm ergs\,s^{-1}}$.  Two regions of diffuse emission, a northern and a southern, are identified with respective emission measures of $\rm{EM} = n_{\rm e}^2 V = 1.5\times10^{54}$ and $1.9\times10^{54}\,{\rm cm^{-3}}$. The attenuating column $N_{\rm H}$ is very low, being $4.1\times10^{20}\,{\rm cm^{-2}}$ for the northern, and $< 10^{20}\,{\rm cm^{-2}}$ for the southern. 

The total mass of the X-ray emitting gas is estimated to be $0.07\,M_{\odot}$, which is roughly $10^{5}$\,yrs of mass-loss of the dominant O5.5 star $\theta^{1}$Ori\,C. The radiative cooling time is estimated to be $\approx 1.8-3.9$\,Myr. The density of the X-ray plasma is determined to be $n_{\rm e} = 0.1-0.5\,{\rm cm^{-3}}$. The X-ray and ionized gas are in approximate pressure equilibrium ($n_{\rm HII} \approx 100\,{\rm cm^{-3}}$), and the hot gas is likely channeled by the cooler denser structures rather than disrupting them by expansion. Leakage of the hot plasma via an X-ray champagne flow into the nearby Eridanus superbubble is suggested.

\subsection{The Rosette Nebula}
The Rosette Nebula is a blister HII region at the tip of the giant Rosette molecular cloud. 
It has a distinct ringlike appearance in both radio and optical images, and is photoionized by the open cluster NGC\,2244 whose stellar winds have cleared a hole in the Nebula's centre \citep{Celnik85,Townsley03}. NGC\,2244 contains 7 O-type stars, all of which have MS luminosity classes, with the earliest spectral type being O4V((f)). A recent analysis of 6 of these stars by \citet{Martins12} determined an upper age limit of 2\,Myr for the most massive stars, in excellent agreement with earlier determinations \citep[e.g.][]{Hensberge00,Park02}. Photometric distance estimates range between 1.4 and 1.7\,kpc, and 1.55\,kpc is adopted in this work. \citet{Wang08} find an absence of mass segregation and conclude that the cluster is not dynamically evolved. The two dominant O stars (HD\,46223, O4V((f)); HD\,46150, O5V((f))z) are widely separated (by at least 3\,pc). In contrast, the O stars in the Trapezium Group and M17 are concentrated within the inner 0.5\,pc. 

\citet{Townsley03} find that soft diffuse X-ray plasma surrounds the OB association and fills the nebula cavity completely. It likely originates from the O-star winds which are thermalized by wind-wind interactions or by shocking against surrounding molecular material.  The X-ray emission is brightest in the central 3\,pc radius, corresponding roughly to the central cavity.  The diffuse emission can be fit by a two-temperature thermal plasma model, with components $kT_{1} = 0.06\pm0.02$ and $kT_{2}=0.8\pm0.1$\,keV and a single absorbing column $N_{\rm H} = 2\pm1 \times 10^{21}\,{\rm cm^{-2}}$. The hotter component is dominant. The intrinsic $0.5-2$\,keV luminosity (for $D=1.55$\,kpc) is $\approx 7\times10^{32}\,{\rm ergs\,s^{-1}}$. There is no significant emission above 2\,keV. Correcting \citeauthor{Townsley03}'s values for a slightly greater assumed distance, the diffuse plasma number density and mass are estimated as $n_{\rm e,x} \sim 0.1\,{\rm cm^{-3}}$ and $M_{\rm x} \sim 0.05\,M_{\odot}$.

\subsection{The Quintuplet Cluster}
The Quintuplet cluster is named after its five brightest stars \citep{Nagata90}. It is located near the Galactic Centre, is unusually dense, and is host to at least 10 massive, windy, WR stars and more than a dozen luminous OB supergiants \citep{Figer99a,Figer99b}. It is somewhat less massive and dense than the Arches cluster, however. Its age is estimated to be about $3.5-4$\,Myr \citep{Figer99b,Liermann12}.

\citet{Law04} determine that the Quintuplet cluster shows thermal diffuse emission with a peak temperature $kT = 2.42 \pm 0.5$\,keV at the cluster centre, and with $L_{\rm x} \sim 10^{34}\,{\rm ergs\,s^{-1}}$.  The diffuse emission is much fainter than that in the Arches and has a very low surface brightness. It also has essentially the same spectral shape as the integrated spectrum from the detected sources. Considering the distance to the cluster, contamination by unresolved point-sources may be an issue.

\citet{Wang06} analyze a deeper \emph{Chandra} exposure. They report the same concerns as \citet{Law04} and in addition note that the extent of the diffuse emission from the Quintuplet cluster is uncertain. With an extraction radius of 1\,arcmin, \citet{Wang06} find that a single-temperature thermal plasma model yields $kT=10^{+4.6}_{-2.7}$\,keV and $N_{\rm H}=3.8^{+0.7}_{-0.5}\times10^{22}\,{\rm cm^{-2}}$, giving an absorption-corrected $2-8$\,keV luminosity of $L_{\rm x} \sim 3 \times 10^{33}\,{\rm ergs\,s^{-1}}$.  The radial diffuse X-ray intensity profile falls off more rapidly than SPH simulations \citep{Rockefeller05} predict.

\subsection{Westerlund\,1}
Westerlund\,1 (hereafter W\,1) is the most massive stellar cluster known in the Galaxy \citep{Clark05,Brandner08}. It contains a rich population of massive stars which include more than 20 WR stars \citep{Crowther06}, more than 80 OB stars, and short-lived transitional objects including luminous blue variables (LBVs), red supergiants (RSGs) and half the currently known population of yellow hypergiants (YHGs) in the Galaxy. Estimates for its age range from $3.6\pm0.7$\,Myr \citep{Brandner08} to $5\pm1$\,Myr \citep{Lim13}. Its distance remains somewhat uncertain, but estimates appear to be converging on the range $4-5$\,kpc \citep[see][and references therein]{Brandner08}. It shows evidence of mass segregation \citep{Lim13}.

At an age of $\sim 4-5$\,Myr, perhaps 100 SNe have already occured in W\,1 \citep[see, e.g., the discussion in][]{Muno06}. The presence of an isolated X-ray pulsar confirms that supernovae have occurred there. However, the likelihood of a recent SNR contributing to the diffuse emission depends on the recent occurrence of a SN event in or near the core, as discussed in \citet{Muno06} and \citet{Kavanagh11}.

The X-ray point sources from a \emph{Chandra} observation are analyzed and reported by \citet{Clark08}, while the diffuse emission is analyzed by \citet{Muno06}. The diffuse emission has an intrinsic ($2-8$\,keV) luminosity of $L_{\rm x} \sim 3\pm1 \times 10^{34}\,{\rm ergs\,s^{-1}}$, and a Lorentzian spatial distribution with a HWHM along the major axis of $25\pm1$ arcseconds ($\sim 0.5$\,pc), and a 5\,arcmin halo. The emission (in the energy range $1.5-8$\,keV) can be fitted with a soft thermal component ($kT_{1} \sim 0.7$\,keV), plus either a harder thermal component ($kT_{2} \sim 3$\,keV) with a low ($\lesssim 0.3$\,solar) Fe abundance, or a nonthermal component with a power-law index $\Gamma \sim 2$. The absorbing column, $N_{\rm H} \sim 2 \times 10^{22}\,{\rm cm^{-2}}$. In the thermal model, $kT_{2}$ increases with distance from the cluster, while in the non-thermal model, $\Gamma$ is significantly steeper in the centre-most region considered. There is no evidence for a recent SN explosion.  Less than $10^{-5}$ of the mechanical luminosity is dissipated as $2-8$\,keV X-rays so it is conjectured that a large fraction escapes into the ISM. However, the X-ray halo between $2-5$\,arcmin ($3-7$\,pc radius) is observed to attain a constant surface brightness of $\approx 7\times10^{-14}\,{\rm ergs\,cm^{-2}\,s^{-1}\,arcmin^{-2}}$, which is not consistent with a cluster wind where almost all of the diffuse X-ray emission is produced within the core radius $R_{\rm c}$ \citep{Stevens03}. A thermal interpretation of the halo of diffuse emission is further challenged by the high temperature and lack of line emission. 

More recently, \citet{Kavanagh11} analyze an XMM-Newton pointing and determine that the hard component in an inner 2\,arcmin radius region is actually thermal, with a clearly detected He-like Fe\,6.4\,keV line. No evidence of a non-thermal component was found.  They report that the diffuse emission has a $2-8$\,keV luminosity of $L_{\rm x} \sim 1.7 \times 10^{33}\,{\rm erg\,s^{-1}}$.

\subsection{The Lagoon Nebula (M8, NGC\,6530)}
The Lagoon Nebula is an HII region associated with the young ($1–3$\,Myr) open cluster NGC\,6530, which contains several O-stars and about 60 B-stars. 
It is about 1.3\,kpc away \citep[see][and references therein]{Henderson12}. Ongoing star formation occurs in several places, notably the Hourglass Nebula (the brightest part of M8) and M8\,E. The Hourglass Nebula is illuminated by an O7V star (Herschel\,36). \citet{Henderson12} argue that NGC\,6530 is slightly younger than the Orion Nebula Cluster (ONC), being $\lesssim1.65$\,Myr assuming the ONC is 2\,Myr old. If the ONC is actually 3\,Myr old, this would give the Hourglass Nebula an age of 2.5\,Myr. The Lagoon Nebula is summarized in \citet{Tothill08}.

\citet{Rauw02} claim that soft diffuse emission was ``probably'' detected from the southern lobe of the Hourglass nebula. The emission can be fitted with an absorbed MEKAL model with $N_{\rm H} = 1.11^{+0.15}_{-0.17} \times 10^{22} \,{\rm cm^{-2}}$, $kT = 0.63^{+0.07}_{-0.05}$\,keV and an intrinsic ($0.2-2.0$\,keV) luminosity of $6.6\times10^{32}\,{\rm ergs\,s^{-1}}$. However, there is unboutedly some contamination from unresolved point sources. No diffuse emission is seen from a qualitative examination of a \emph{Chandra} observation of M8 which did not cover the Hourglass Nebula \citep{Townsley03} - see also \citet{Damiani04}.

\subsection{The Arches Cluster}
The Arches cluster, like the Quintuplet cluster, lies close to the Galactic Centre, being only 26\,pc away in projection \citep[see, e.g.][]{Figer99b}. It is slightly younger \citep[$2-2.5$\,Myrs,][]{Najarro04,Martins08} and more massive \citep{Figer02} than the Quintuplet cluster.
\citet{Clarkson12} have recently measured the kinematic mass of the cluster using \emph{Keck}-LGS adaptive optics.

The deepest \emph{Chandra} observation to date is by \citet{Wang06}.
Diffuse thermal X-ray emission with a prominent Fe K$\alpha$ 6.7\,keV emission line is seen from the core of the Arches cluster. The surface intensity declines steeply with radius, consistent with a cluster wind origin.
This central ($r < 0.6$\,pc) "plume" region can be fitted with a single temperature NEI plasma model with $kT=2.56$\,keV, $\tau = 1.2\times10^{11}\,{\rm cm^{-3}\,s}$, $N_{\rm H}=1.1\times10^{23}\,{\rm cm^{-2}}$ and an intrinsic (2-8keV) luminosity of $L_{\rm x} = 3.8\times10^{33}\,{\rm ergs\,s^{-1}}$.

In contrast, the emission in the outer regions of the cluster shows a prominent line at 6.4\,keV, a power-law continuum emission of non-thermal origin, and a non-axissymmetric spatial distribution with a bowshock morphology.  This may result from an ongoing collision between the cluster and the adjacent molecular cloud, which has a relative velocity of $120\,{\rm km\,s^{-1}}$. The interpretation of the 6.4\,keV Fe K$\alpha$ flourescence emission from neutral Fe is still debated, with \citet{Wang06} favouring a cosmic ray origin but \citet{Capelli11} favouring photoionization of nearby molecular clouds by X-ray photons.  \citet{Wang06} find that the SE extension (which is where the 6.4\,keV emission is) is best fitted with a PL+Gaussian spectral model with $\Gamma = 1.3^{+1.4}_{-1.1}$ and $N_{\rm H}=6.2\times10^{22}\,{\rm cm^{-2}}$, and
has an intrinsic ($2-8$\,keV) luminosity $L_{\rm x} = 4.1\times10^{33}\,{\rm ergs\,s^{-1}}$. An even more extended ``LSBXE'' region is fitted with
a MEKAL+PL+GAUSSIAN spectral model with $kT=0.45$\,keV, $\Gamma=1.3$(fixed) and $N_{\rm H}=9.2\times10^{22}\,{\rm cm^{-2}}$, with an intrinsic ($2-8$\,keV) $L_{\rm x} = 1.2\times10^{34}\,{\rm ergs\,s^{-1}}$. 

Fig.~15 in \citet{Wang06} shows the radial diffuse X-ray intensity profiles around the Arches cluster. It falls off much less rapidly than simulations \citep{Rockefeller05}.

\subsection{NGC\,3576 (RCW\,57)}
NGC\,3576 is a giant HII region located at a distance of $2.8\pm0.3$\,kpc \citep{Figueredo02}, and which is projected to within 30 arcminutes of the more distant region NGC\,3603. Together these regions make up the RCW\,57 complex. The ionizing cluster for NGC\,3576 remains deeply embedded in the centre of an extended filamentary dust cloud, and not enough massive stars have yet been found to account for the radio luminosity \citep{Figueredo02,Barbosa03}.

The evidence for sequential star formation in NGC\,3576 remains controversial \citep[see][and references therein]{Purcell09}. These authors provide a schematic of the region (their Fig.~16).

\citet{Townsley11} analyzed two \emph{Chandra} pointings. A southern pointing was centered on NGC\,3576, while a northern pointing was designed to search for a young stellar cluster associated with the O8V+O8V eclipsing binary HD\,97484 (EM\,Car) and the O9.5Ib star HD\,97319. Diffuse emission is seen to the SE of NGC\,3576 (hereafter identified as NGC\,3576S), while hard X-rays were seen in the northern pointing. \citet{Townsley11} identified these sources as NGC\,3576S and NGC\,3576N, respectively. The northern pointing revealed a young cluster (termed NGC\,3576OB) which appears older than NGC\,3576 to the south.

NGC\,3576S requires a 2-temperature spectral fit, with $kT_{1}=0.31^{+0.06}_{-0.07}$ and $kT_{2}=0.53$. The absorbing columns are $N_{\rm H}=1.3\times10^{22}\,{\rm cm^{-2}}$ and $N_{\rm H}=2.5\times10^{21}\,{\rm cm^{-2}}$, respectively. The softer component dominates the total emission which has an intrinsic $L_{\rm x}=1.1\times10^{34}\,{\rm ergs\,s^{-1}}$. \citet{Townsley11} suggests that the hot plasma responsible for this emission has forced itself out through a low-density pathway, analogous to the outflow seen from M17, but seen more face-on and at a slightly earlier phase.  A gaussian at 0.72\,keV (which accounts for 16\% of the total emission) is required for a good fit. This may represent charge exchange processes.

In contrast NGC\,3576N requires the presence of a power-law component in spectral models. The intrinsic luminosity is $L_{\rm x}=1.2\times10^{34}\,{\rm ergs\,s^{-1}}$, 24\% of which is contributed by a power-law continuum. A 3-temperature model is also required for a good fit, with the hardest NEI component ($kT_{3}=0.7$\,keV) accounting for 48\% of the total emission. \citet{Townsley11} speculate that the diffuse emission from this region has been enhanced by a recent cavity SN. There is no evidence for charge exchange.

\subsection{NGC\,3603}
The luminous giant HII region NGC\,3603 contains the compact star cluster HD\,97950, which is one of the most massive young star clusters in the Milky Way. It contains 3 core H-burning WN-stars and up to 50 O-stars \citep{Drissen95}. The most massive stars in the core appear to be coeval with an age of about 1\,Myr, while less massive stars and stars in the cluster outskirts appear to be older \citep[][and references therein]{Melena08,Pang13}. It shows clear mass segregation, despite its young age. \citet{Pang13} suggest that dynamical processes may have been dominant for the high mass stars. Star formation appears to have occurred almost instantaneously, with \citet{Kudryavtseva12} deriving an upper limit to the age spread of 0.1\,Myr. The distance to NGC\,3603 is thought to be $7\pm1$\,kpc \citep{Harayama08}. \citet{Banerjee13} explore whether a phase of substantial gas-expulsion has occurred in NGC\,3603.
 
A \emph{Chandra} cycle 1 observation was presented by \citet{Moffat02}, who noted diffuse X-ray emission within a central region of 2 arcmin radius with an intrinsic luminosity $L_{\rm x}=2\times10^{34}\,{\rm ergs\,s^{-1}}$. However, this is 20\% of the integrated point source emission within this region and may be completely due to undetected point sources.

\citet{Townsley11} recently re-analyzed this observation, finding 1328 point sources compared to the 348 sources found by \citet{Moffat02}. The diffuse X-ray emission is anti-coincident with the mid-IR emission which traces the surrounding heated dust. This is consistent with the hot plasma from the shocked stellar winds filling the cavities that they have carved. Excluding an area around the core of NGC\,3603 (which is likely dominated by unresolved point sources) and a region to the west (which may contain foreground emission related to the NGC\,3576 cluster), the diffuse X-ray emission is dominated by an NEI thermal plasma component with $kT_{1} = 0.53$\,keV, $\tau=2\times10^{10}\,{\rm cm^{-3}\,s}$, $N_{\rm H}=2\times10^{22}\,{\rm cm^{-2}}$, and which contributes 86\% of the total intrinsic $L_{\rm x}=2.6\times10^{35}\,{\rm ergs\,s^{-1}}$. No evidence for charge exchange processes was found though the exposure is quite short.

\subsection{NGC\,2024 (The Flame Nebula)}
The Flame Nebula, NGC\,2024, is one of the nearest sites of massive star formation \citep[$D=415$\,pc,][]{Anthony-Twarog82}. It is part of the Orion\,B giant molecular cloud \citep[e.g.][]{Mitchell01} and is near the Horsehead Nebula. A 3D structure of the region was proposed by \citet{Barnes89} \citep[see also][]{Emprechtinger09}.  \citet{Bik03} suggested that the O8V-B2V star IRS\,2b is the ionizing source of the HII region, but \citet{Burgh12} note that it could be a supergiant. The age of NGC\,2024 is unclear, with estimates ranging from 0.3\,Myr \citep{Meyer96} to several Myr \citep{Comeron96}.

Diffuse X-ray emission with a radius of 0.5\,pc from the centre of NGC\,2024 was reported by \citet{Ezoe06b}. The emission has a very
hard continuum ($kT > 8$\,keV) and shows a He-like Fe\,K$\alpha$ line. Fitting the data with a ``leaky absorber'' model (where emission from a single temperature plasma reaches the observer via two paths with different absorption) returns $kT \approx 11$\,keV with $N_{\rm H}=0.21 \times10^{22}$ and $3.3\times10^{22}\,{\rm cm^{-2}}$. The intrinsic X-ray luminosity in the $0.5-7$\,keV band is $L_{\rm x}=2\times10^{31}\,{\rm ergs\,s^{-1}}$. \citet{Ezoe06b} note that a single massive star with a wind comparable to, or stronger than, that of a typical B0.5V star has enough energetics to power the
observed X-ray emission. This work shows that diffuse emission is present in a MSFR in which only late O to early B stars exist.



\label{lastpage}

\end{document}